\documentclass[12pt]{elsart}
 \usepackage{graphics}
 \begin{document}
\begin{frontmatter}
 %%% article title
 \title
 {Low temperature resonances in the electron heat capacity and its spectral distribution in finite systems.}
 \author[Radium]{N.K. Kuzmenko}, %\corauthref{cor}
 %\corauth[cor]{Corresponding author.} \ead{kuzmenko@NK9433.spb.edu}
 \author[Univ]{V.M. Mikhajlov}
 \address[Radium]{V.G.Khlopin Radium Institute, 194021
 St.-Petersburg, Russia}
 \address[Univ]{Institute of Physics St.--Petersburg State
 University 198904, Russia}
% \date{\today}
 %%% abstract
\begin{abstract}
Temperature variations of the heat capacity ($C$) are studied in a low temperature regime $T<\delta_F\sim\varepsilon_F/N$  for $2D$-, and $3D$-systems with $N\sim10^2\div 10^4$ treated as a canonical ensemble  of $N$-noninteracting fermions. The analysis of $C$ is performed  by introducing function $\varphi (\varepsilon )$, the spectral distribution of $C$, that gives the contribution of each single-particle state to $C$. This function has two peaks divided by the energy interval $\Delta\varepsilon\approx (2\div 5)T$. If at some temperature $T_{res}$ there takes place a resonance i.e. the positions of these peaks coincide with energies of two levels  nearest to $\varepsilon_F$ then $C$ vs $T$ can show a local maximum  at $T_{res}$. This gives possibility to assess the single-particle level spacings near the Fermi level.

\end{abstract}
\end{frontmatter}

\section{Introduction}

The heat capacity of solids characterizes the temperature variation of the internal energy corresponding to vibrations of lattice ions and movement of so called free electrons confined in solid. At temperatures $T$ above the Debye temperature $T_D$ the phonon heat capacity prevails, fall in temperature below $T_D$ first equalizes the phonon and electron contributions to $C$ and then at $T\ll T_D$ the main part of $C$ belongs to electrons~\cite{ashcroft}. This general tendency in temperature evolution of $C$ is observed in both macroscopic and mesoscopic systems though in the latter finite sizes can reveal themselves in the temperature and size dependence of $T_D$ ~\cite{gu}. Inside the range, $0<T<T_D$, the relationship between the electron and phonon components is regulated by the particle number $N$, the ratio of $T_D$ to the electron Fermi energy $\varepsilon_F$ (both quantities are material dependent). In mesoscopic systems the spatial shape and symmetry of a sample can also affect this relationship. However the phonon and electron contributions to $C$ are practically independent that allows us to study here the low temperature electron heat capacity of mesoscopic systems separately from the phonon one.

The electron movement in normal (nonsuperconducting) solids at low temperatures can be described in the framework of the NONINTERACTING PARTICLE MODEL
($NPM$) provided particles are understood as quasiparticles introduced by the Fermi liquid theory~\cite{abrikosov}. This theory treats quasiparticles as compound fermion type excitations with an effective mass $m^*$ that differs from the free electron mass $m$ and material dependence of $m^*$ causes the  difference in the Fermi energies $\varepsilon_F$ of a solid and the Fermi gas in the same volume. These quantities, $m^*$ or $\varepsilon_F$, via $\rho_0(\varepsilon_F)$, the electron state density in the vicinity $\varepsilon_F$, determine the prefactor in the linear in $T$ Sommerfeld's law for the electron heat capacity
\begin{equation}\label{Clin1}
C/k_B=\frac{\pi^2}{3}\rho_0(\varepsilon_F)T,
\end{equation}
in Eq.~(\ref{Clin1}) and hereinafter $T$ is measured in energy units. In macroscopic systems this law is fulfilled in a wide low temperature range, $0\leq T\ll\varepsilon_F$, irrespective of the size and shape of the body. For some materials with $m^*\gg m$ there are experimental and theoretical evidences pointing out that the only parameter $m^*$ is insufficient to describe $C$ in a wide temperature range since the heat capacity as a function of $T$ can include a kink  separating two linear temperature dependencies with different slopes~\cite{toshi}. The corresponding temperature is relatively high and we suppose that such radical change in $m^*$ does not concern a mesoscopic system at low temperature. In finite systems Eq.~(\ref{Clin1}) is applicable down to an interval of $T$ of order of interlevel spacings ($\sim \varepsilon_F /N$). Within it $C$ increases exponentially and can display some local extrema which we shall consider in detail below. In mesoscopic systems there is also a transitional region of temperature quasilinearity where $\rho_0(\varepsilon_F)$ depends on the shape of the body though these corrections are small $\sim N^{-1}$ and $\sim N^{-2}$~\cite{balian,bohr}.

The temperature scale in finite fermion systems explicitly depends on $\varepsilon_F$ that is material dependent: for electrons in metals $\varepsilon_F\sim 10^4\div 10^5$ $K$, in heterostructures $\sim 10^2$ $K$ and for trapped Fermi gases $\varepsilon_F\sim 1\mu K$. Therefore, not to be bound to an individual mesoscopic system we measure all energies and the temperature in $\delta_F$, the average level spacing near $\varepsilon_F$
\begin{equation}\label{deltaF0}
\delta_F=\widetilde{d_F}/\rho_0(\varepsilon_F)\sim\varepsilon_F/N,
\end{equation}
$\widetilde{d_F}$ is the average degeneracy near $\varepsilon_F$, for asymmetric systems as a rule $\widetilde{d_F}=2$, for symmetric ones values  $\widetilde{d_F}>2$ will be specially pointed out below. Thus all results of our calculations in Secs. 3,4 can be projected on any real system if its $\varepsilon_F$ is known. Owing to Eq.~(\ref{deltaF0}) the term ``low temperatures'' acquires a more definite meaning. In the first instance it will be used for the temperature range smaller than or of order of $\delta_F$ but these temperatures have to be really low, much less than $T_D$ otherwise the electron heat capacity cannot be observed.

In the finite Fermi systems~\cite{migdal} besides the effective mass (we suppose that $m^*$ is of the same order as in macroscopic bodies) the quasiparticle description makes use of several additional parameters pertaining to the quasiparticle mean field (the depth, radius, diffusivity of the potential and others). These parameters are formed mainly by the ion surroundings and among real mesoscopic systems there are cases of potentials close to rectangular $3D$-wells as it occurs in metal clusters~\cite{heer} whereas in quantum dots the best description of electron shells is given by $2D$-oscillators~\cite{kouwen}. These cases will be used in illustrative examples of Secs.3,4. Such description idealizes and simplifies the real picture of the electron movement in mesoscopic systems as it neglects effects of disorder, Coulomb and phonon interactions. However we believe that detailed ascertainment of the mean field role can help to reveal the impact of the other factors on measured thermodynamic quantities.

Independently of the type of electron confinement and of that whether a system is integrable or not the common feature of quantum energy spectra in different finite systems consists in the appearance of such irregularities in the eigenvalue distributions as level bunchings or high degenerated levels in system possessing some symmetry. These irregularities can reveal themselves in measured quantities e.g. in the low temperature electron heat capacity. This fact was for the first time established by Fr\"{o}lich~\cite{frolich} for a model spectrum with high degenerated levels. Later an analogous result was obtained for spherical alkali clusters by Brack et al~\cite{brack} where the single-particle potential was found by the temperature Hartree-Fock method. In both cases the low temperature variation of the heat capacity displays local maxima at those particle numbers which correspond to completely occupied high degenerated levels. On the other hand the calculations of Denton et al~\cite{denton} showed that a uniform single-particle spectrum with spin degenerated levels and equal spacings between them practically gives a monotonous increase of $C$ if the particle number conservation is exactly taken into account. The comparison of these results has stimulated us to look for the cause of such different behavior of $C$ vs $T$ and the investigation of this problem is the subject of the following sections.

Our approach to this problem supposes that each mesoscopic system possesses a quite definite shape, linear size and particle number. Now such systems  can be fabricated by using the modern technologies and  examples of such possibilities are described in Refs.~\cite{kouwen,delft,prinz,yanson,diaz,mares}. Thus our description of mesoscopic systems follows the line of Refs.~\cite{frolich,brack} but differs from the approach developed by Kubo~\cite{kubo,perenboom,halperin,nagaev}. This approach takes into account that in many cases shapes, sizes and particle numbers of mesoscopic systems cannot be fixed exactly and so thermodynamic quantities of such systems have to be calculated with single -particle spectra defined by means of statistical distributions. However even in that case when experimental conditions make it possible to study only a set of systems parameters of which are given in some limits our results can be employed through an averaging procedure.

The particle number conservation plays an important role in low temperature investigations of a mesoscopic system. As shown in Refs.~\cite{brack,denton} the results of the canonical and grand canonical calculations can noticeably differ in this temperature range. In Ref.~\cite{kuzmenko} we have developed a canonical polynomial method appropriate for studying finite systems in wide ranges of $T$ and $N$. Here the method is modified to give a convenient qualitative description of the heat capacity and show the resonance character of amplifying the heat capacity at some temperatures in the low temperature range.

The subjects mentioned above are distributed over the following sections. In Sec.2 we introduce a new representation for the canonical occupation numbers as a continuous function of energy and then on this basis we construct the spectral heat capacity $\varphi(\varepsilon )$. The latter is especially  appropriate to analyze the influence of the irregularities in the single-particle spectrum on $C$ in $NPM$ that allows us to explain in Sec.3 the cause of arising the low temperature maxima in $C$. In this section several examples of such maxima are given for both asymmetric and symmetric systems. In Sec.4 we consider variations of $C$ caused by the shape variations in the $2D$-, and $3D$-mesoscopic systems at an invariable area or volume i.e. at the particle number conservation. Conclusions are given in Sec.5.

\section{The spectral heat capacity distribution}

Thermodynamic properties of the CANONICAL ENSEMBLE ($CE$) of $N$ independent fermions are traditionally treated in the framework of the approximate method which virtually proceeds from the assumption that $CE$ occupation numbers ($n_s$) of single-particle states (their energies and degeneracies are $\varepsilon_s$ and $d_s$ respectively) are the same as in the GRAND CANONICAL ENSEMBLE ($GCE$) i.e. $n_s=f_s$ where $f_s$ is the Fermi-Dirac function
\begin{equation}\label{FermiDir}
f_s=\left [ 1 + e^{\beta(\varepsilon_s-\lambda)}\right ]^{-1};\;\;\;\; \beta=T^{-1}.
\end{equation}
In Eq.~(\ref{FermiDir}) $\lambda$ is the chemical potential chosen so that the ensemble averaged particle number at each $T$ is equal to a fixed value of $N$ in a studied system
\begin{equation}\label{eqNf}
N=\int_0^{\infty}\rho(\varepsilon)f(\varepsilon)d\varepsilon.
\end{equation}
In Eq.~(\ref{eqNf}) and hereafter we suppose that single-particle energies $\varepsilon\ge 0$ and $\rho(\varepsilon)$ is the level density
\begin{equation}\label{rhoex}
\rho(\varepsilon)=\sum_{s}\delta(\varepsilon-\varepsilon_s)d_s,
\end{equation}
which depends on $N$ via $\varepsilon_s$ due to the connection between the size and particle number in mesoscopic systems.

In the exact canonical description the occupation numbers $n_s$ are such that the equation analogous with Eq.~(\ref{eqNf}) is provided automatically
\begin{equation}\label{eqNn}
N=\int_0^{\infty}\rho(\varepsilon)n(\varepsilon)d\varepsilon.
\end{equation}
irrespective of the temperature and of the energy count point whose role in $GCE$ is played by $\lambda$.

$T$ and $\lambda$ are treated as independent variables in the classic definition of $GCE$ while in $CE$ those are $T$ and $N$ (in both cases the volume $V\sim N$ keeps fixed therefore below the values of the heat capacity $C$ will be considered only at this condition, $C\equiv C_V$). In order to stress the temperature dependence of $\lambda$ at the description with $n_s=f_s$ this approximate method will be designated by the term the EQUIVALENT $GCE$ ($EGCE)$ ~\cite{goldstein}.

In the $EGCE$ formalism the partition function $Z_{EGCE}$ can be constructed so as to describe occupation numbers with help of $f_s$ and simultaneously reproduce the correct relation between the free energy $\mathcal{F}_{EGCE}$ and the thermodynamical potential $\Omega$
\begin{eqnarray}\label{ZetGCE}                                       %(5)
\Omega=-T\ln Z_{GCE}; \;\;\; Z_{GCE}=\prod_s\{ 1+exp\left[-\beta(\varepsilon_s-\lambda)\right]\}^{d_s},\\
\mathcal{F}_{EGCE}=\lambda N+\Omega=-T\ln Z_{EGCE} \label{ZetEGCE}
\end{eqnarray}
Eqs.~(\ref{ZetGCE}) and (\ref{ZetEGCE}) can be consistent if
\begin{equation}\label{ZZ}
Z_{EGCE}=exp(-\beta\lambda N)Z_{GCE}
\end{equation}
that immediately leads to $(n_s)_{EGCE}=f_s$.

Therefore in the $EGCE$ description the internal energy $E_{EGCE}$ and heat capacity $C_{EGCE}$ are represented in the standard form
\begin{eqnarray}\label{Eegce}
E_{EGCE}=-\frac{\partial}{\partial\beta}\ln Z_{EGCE}=\int_0^{\infty}\rho(\varepsilon)\varepsilon f(\varepsilon)d\varepsilon,\\
C_{EGCE}/k_B=\frac{\partial E_{EGCE}}{\partial T}=\beta^2\frac{\partial^2}{\partial\beta^2}\ln Z_{EGCE}=\beta^2\int_0^{\infty}\rho(\varepsilon)\varepsilon\left[-\frac{\partial f(\varepsilon)}{\partial\beta}\right]d\varepsilon.\label{CZegce}
\end{eqnarray}
The temperature variations of $\lambda$ in the $EGCE$ theory explicitly manifest themselves in $\partial f/\partial\beta$ which depends on $\beta\partial\lambda/\partial\beta$
\begin{eqnarray}\label{dfdb}
\frac{\partial f(\varepsilon)}{\partial\beta}=\left(\varepsilon-\lambda-\beta\frac{\partial\lambda}{\partial\beta}\right)
\frac{\partial f(\varepsilon)}{\beta\partial\varepsilon}.
\end{eqnarray}
The quantity $\beta\partial\lambda/\partial\beta$ is straightforwardly calculated allowing for the temperature independence of $N$ in the $EGCE$ theory \begin{equation}\label{dNdb}                                                 %(11)
\frac{\partial N}{\partial\beta}=0=\left(\lambda+\beta\frac{\partial\lambda}{\partial\beta}\right)\int_0^{\infty}\rho(\varepsilon)\frac{\partial f(\varepsilon)}{\beta\partial\varepsilon}d\varepsilon-\int_0^{\infty}\rho(\varepsilon)\varepsilon\frac{\partial f(\varepsilon)}{\beta\partial\varepsilon}d\varepsilon.
\end{equation}
The first integral in this equation is the particle number dispersion in both $GCE$ and $EGCE$
\begin{equation}\label{dispersion}                                            % (12)
\int_0^{\infty}\rho(\varepsilon)\left[-\frac{\partial f(\varepsilon)}{\beta\partial\varepsilon}\right]d\varepsilon=\langle \hat{N}^2\rangle -\left(\langle \hat{N}\rangle\right)^2
\end{equation}
$\langle \hat{N}^2\rangle$ and $\left(\langle \hat{N}\rangle\right)^2$ are the ensemble averages of operators $\hat{N}^2$ and $\hat{N}$. The quantity $(\lambda+\beta\partial\lambda/\partial\beta)$ defines also values of $\partial E_{EGCE}/\partial N$, if $N$ and $T$ are high enough and $\partial E/\partial N$ makes sense (at small variations of $N$ the single-particle spectrum is fixed for an $N$ particle system on account of the volume conservation)
\begin{equation}\label{dEdN}                                                  %(13)
\frac{\partial E_{EGCE}}{\partial N}=\lambda+\beta\frac{\partial\lambda}{\partial\beta}
\end{equation}
Hence one can obtain that ($S$ is the entropy)
\begin{equation}\label{entropy}
\frac{\partial \mathcal{F}_{EGCE}}{\partial N}=\lambda; \;\;\;\;\;\; \frac{\partial S}{\partial N}=-\frac{\partial\lambda}{\partial T}
\end{equation}
In consequence of Eqs.~(\ref{dfdb}) and (\ref{dNdb}) the integral for $C_{EGCE}$ can be represented in such form that the integrand is obviously positive
\begin{eqnarray}\label{Cegce}                                                 %(15)
C_{EGCE}=\beta^2\int_0^{\infty}\rho(\varepsilon)\varepsilon\left(\varepsilon-\lambda-\beta\frac{\partial\lambda}{\partial\beta}\right)
\left[-\frac{\partial f(\varepsilon)}{\beta\partial\varepsilon}\right]d\varepsilon=\nonumber\\
\beta^2\int_0^{\infty}\rho(\varepsilon)\left(\varepsilon-\lambda-\beta\frac{\partial\lambda}{\partial\beta}\right)^2
f(\varepsilon)\left(1- f(\varepsilon)\right)d\varepsilon .
\end{eqnarray}
Below to emphasize the role of the term $\beta\partial\lambda/\partial\beta$, we adduce results obtained without this term ($\beta\partial\lambda/\partial\beta=0$). Such results will be denoted by letters $GCE$.

The exact canonical partition function in the polynomial representation was suggested in Ref.~\cite{kuzmenko}. Here, to have a uniform description for $EGCE$ and $CE$ we introduce a factor $exp\left(-\beta\lambda_cN\right)$ into the partition function of Ref.~\cite{kuzmenko}:
\begin{equation}\label{Zce}             %(16a)
Z_{CE}\equiv Z(N)=\left[ N \right] exp\left(-\beta\lambda_c N\right)
\end{equation}
where designations $\left[ N \right]$ and $\left[ n \right]$, that appears below, are employed for symmetric polynomials the orders of which are $N$ and $n$ respectively. These polynomials are defined in a space of variables $q_s$
\begin{equation}\label{qs}
q_s=exp\{-\beta(\varepsilon_s-\lambda_c)\},                 % (17)
\end{equation}
e.g.
\begin{eqnarray}
\left [1\right ]=\sum_sd_sq_s; \;\;\;\; \left [2\right ]=\sum_{s_1<s_2}d_{s_1}q_{s_1}d_{s_2}q_{s_2}+\frac{1}{2}\sum_{s}d_s(d_s-1)q_s^2.\nonumber
\end{eqnarray}
Recurrent relations for polynomials and convergence of series for $\left[ N \right]$ are considered in Ref.~\cite{kuzmenko}.

The auxiliary parameter $\lambda_c$ in Eqs.~(\ref{Zce}), (\ref{qs}) separates the energy regions above and below the Fermi energy $\varepsilon_F$ and serves as a point to count single-particle level energies. However, $Z(N)$ in Eq.~(\ref{Zce}) does not comprise $\lambda_c$ at all and this independence from $\lambda_c$ is inherent in all functions generated by $Z(N)$, i.e.
\begin{equation}\label{ZN1}
Z(N)=\left[ N \right] \mid_{\lambda_c =0}.
\end{equation}
As shown in Ref.~\cite{kuzmenko} the polynomial $\left[ N \right]$ can be expanded in a rapidly convergent series in products of $n$ order polynomials, $n<N$. In these series each term corresponds to $n$-particle-$n$-hole excitations with respect to the ground state at $T=0$ in which all single-particle states are filled up to and including the Fermi level ($F$). The simplest example of such expansion is $Z(N)$ for a system with $n_F=d_F$ ($n_F$ is the occupation number of the Fermi level at $T=0$).
\begin{eqnarray}\label{ZN2}
Z(N)=Z_0(N)\widetilde{Z}(N),\nonumber\\
Z_0(N) = exp\left(-\beta E_0(N)\right);\;\;\;\;  E_0(N)=\sum_{s\le F}d_s\varepsilon_s,\nonumber\\
\widetilde{Z}(N)=1+\sum_{n=1}^N [n]\overline{[n]},
\end{eqnarray}
$[n]$, $\overline{[n]}$ are symmetric $n$-th order polynomials. The particle polynomial $[n]$ is composed of $q_s$ $(\varepsilon_s>\lambda_c)$, Eq.~(\ref{ZN2}),the hole polynomial of $q_s^{-1}$ $(\varepsilon_s<\lambda_c)$,
\begin{equation}\label{eFlamb}
\varepsilon_F<\lambda_c<\varepsilon_{F+1},
\end{equation}
$F+1$ or in the general case $F+k$ and $F-k$ are the first or $k$-th level above and under $F$ respectively. If $F$ is incompletely occupied at $T=0$ or there is a necessity to count energies not from $\lambda_c$ given by Eq.~(\ref{eFlamb}) but from $\varepsilon_{F+i}$ or $\varepsilon_{F-k}$ then single-particle energies have to be summed in Eq.~(\ref{ZN2}) up to $(F+i)$-th or only down to $(F-k)$-th level whereas the series in particle and hole polynomials have to begin now with $[0]\overline{[\mu]}$ and $[\nu]\overline{[0]}$ respectively, $[0]=\overline{[0]}=1$;  $\mu=\sum_{s\le F+i}d_s -N$; $\nu=N-\sum_{F-k}d_s$. By using this possibility one can write down the partition functions of $N\pm \nu$ particles as following
\begin{eqnarray} \label{ZNpm}
 Z(N\pm\nu)=\exp\{-\beta\left [E_0(N)\pm\lambda_c(N)\nu\right]\}\left \{ \begin{array}{ll}
\sum_{n=0}[\nu +n]\overline{[n]}\\
\sum_{n=0}[n]\overline{[\nu+n]},
\end{array}
\right.
\end{eqnarray}
where $\lambda_c(N)$ is chosen so that $\lambda_c>\varepsilon_F(N)$ i.e. near $\varepsilon_F$ for the $N$-particle system and $E_0(N)=\sum_{s\le F(N)}d_s\varepsilon_s$, $F(N)$ is the Fermi level of $N$ particles.

To find the canonical occupation number $n_s$
\begin{equation}\label{ns}
n_s=-\frac{T}{d_s}\frac{\partial}{\partial\varepsilon_s}ln Z(N)=\frac{q_s}{d_s}\frac{\partial}{\partial q_s}ln Z(N)
\end{equation}

($n_s$ characterizes the occupation probability of level $s$ for only one electron, the full number of particles in state $s$ is $n_sd_s$)
\begin{equation}\label{Nsum}                                                 %(21b)
N=\int_0^{\infty}\rho(\varepsilon)n(\varepsilon)d\varepsilon=\sum_sd_sn_s
\end{equation}
we take advantage of properties of symmetric polynomials
\begin{eqnarray}\label{sympol1}
\frac{q_s}{d_s}\frac{\partial}{\partial q_s}\left[ N\right]=\left[ N\right]-\sum_{m=1}^{M-N}(-)^{m+1}\left[ N+m\right]q_s^{-m}=\\
=\sum_{m=1}^{N}(-)^{m+1}\left[ N-m\right]q_s^{m}\label{sympol2}
\end{eqnarray}
In Eq.~(\ref{sympol1}) $M$ is the maximum polynomial order for the single-particle basis we take into account, as a rule $M\gg N$. Both Eqs.~(\ref{sympol1}) and (\ref{sympol2}) are exact. Thus any of them can be used to determine $n_s$. Nevertheless the convergence of the series can be hastened if Eq.~(\ref{sympol1}) is applied for states with $\varepsilon\le\varepsilon_F$ $(q_s^{-1}<1)$ and Eq.~(\ref{sympol2}) for states with $\varepsilon>\varepsilon_F$ $(q_s<1)$
\begin{eqnarray}
n(\varepsilon\le\varepsilon_F)=1-\sum_{m=1}^{M-N}(-)^{m+1}\exp [\beta m(\varepsilon-\lambda_{m+})],\label{nelt}\\
n(\varepsilon >\varepsilon_F)=\sum_{m=1}^{N}(-)^{m+1}\exp [-\beta m(\varepsilon-\lambda_{m-})],\label{ngt}\\
\exp [\mp\beta m\lambda_{m\pm}]=Z(N\pm m)/Z(N),\label{expm}\\
m\lambda_{m\pm}=\pm\left [\mathcal{F}(N\pm m)-\mathcal{F}(N)\right].\label{mlambda}
\end{eqnarray}
Eqs.~(\ref{expm}) and (\ref{mlambda}) indicate that each power $m$ in Eqs.~(\ref{nelt}), (\ref{ngt}) enters into $n(\varepsilon)$ with its own chemical potential not arbitrary chosen but conditioned by the difference of the canonical free energies, Eq.~(\ref{mlambda}), which have to be calculated at invariable volume, i.e. with the same single-particle energy spectrum as in the case of $N$ particles.

The comparison of Eqs.~(\ref{nelt}) and (\ref{ngt}) with similar expansion of $f(\varepsilon)$
\begin{eqnarray}\label{felt}
f(\varepsilon\le\varepsilon_F)=1-\sum_{m=1}^{\infty}(-)^{m+1}\exp [\beta m(\varepsilon-\lambda)]\\
f(\varepsilon >\varepsilon_F)=\sum_{m=1}^{\infty}(-)^{m+1}\exp [-\beta m(\varepsilon-\lambda)]\label{fgt}
\end{eqnarray}
shows that the distinction between $n(\varepsilon)$ and $f(\varepsilon)$ is caused in general by the difference between $\lambda_{m\pm}$ in  Eqs.~(\ref{nelt}), (\ref{ngt}) and the $GCE$ chemical potential $\lambda$, though at high enough $N$ and $T$ ( the case considered in Eq.~(\ref{entropy})) they become indistinguishable at least for small $m\ll N$ when

\begin{displaymath}
\mathcal{F}(N\pm m)\simeq\mathcal{F}(N)\pm m\frac{\partial\mathcal{F}(N)}{\partial N};\;\;\;\;
\lambda_{m\pm}\longrightarrow\frac{\partial\mathcal{F}(N)}{\partial N}.
\end{displaymath}
The difference between $\lambda$ and $\lambda_{m\pm}$ can be assessed through the relation of $Z_{EGCE}$ and $Z_{CE}$ obtained by expanding $Z_{GCE}$, Eq.~(\ref{ZetGCE}), in symmetrical polynomials
\begin{equation}\label{ZGCEpol}                                                 %(27)
Z_{GCE}=\left[N\right]_{(\lambda)}+\sum_{m=1}^{N}\left[N-m\right]_{(\lambda)}+\sum_{m=1}^{M-N}\left[N+m\right]_{(\lambda)}.
\end{equation}
The subscript $(\lambda)$ is applied in Eq.~(\ref{ZGCEpol}) to point out that polynomials $\left[N\pm m\right]_{(\lambda)}$ are composed of $q_s$, Eq.~(\ref{qs}), in which $\lambda_c$ is replaced by the $GCE$ potential $\lambda$. Hence, $Z_{GCE}$ gains the form including potentials $\lambda_{m\pm}$ defined by Eqs.~(\ref{expm}), (\ref{mlambda}):
\begin{eqnarray}
Z_{EGCE}=e^{-\beta\lambda N}Z_{GCE}=Z(N)R(N);\nonumber\\
R(N)=1+\sum_{m=1}^{N}\frac{Z(N-m)}{Z(N)}e^{-\beta\lambda m}+\sum_{m=1}^{M-N}\frac{Z(N+m)}{Z(N)}e^{\beta\lambda m}=\nonumber\\
1+\sum_{m=1}^{N} e^{-\beta m(\lambda-\lambda_{m-})}+\sum_{m=1}^{M-N}e^{-\beta m(\lambda_{m+}-\lambda)}.\label{ZEGCEpol}
\end{eqnarray}
As shown in Eqs.~(\ref{Zce}), (\ref{ZN1}), $Z(N\pm m)\equiv Z_{CE}(N\pm m)$ does not comprise $\lambda$ or $\lambda_c$. Thus, independently of values $\lambda$ and $\lambda_{m\pm}$ the quantity $\ln R(N)>0$ and
\begin{equation}\label{FEGCEle}
\mathcal{F}_{EGCE}<\mathcal{F}_{CE}
\end{equation}
At $T\longrightarrow0$ the $EGCE$ and $CE$ free energies are identical. Consequently Eq.~(\ref{ZEGCEpol}) for superlow temperatures gives an estimation:
\begin{equation}\label{lambdaest}                                                 %(29)
\lambda_{m+}\ge\lambda\ge\lambda_{m-}.
\end{equation}
In another temperature regime, $\delta\ll T\ll\varepsilon_F$ ($\delta$ is the mean level spacing), the ratio $Z_{EGCE}/Z_{CE}\sim(T/\delta)^{1/2}$, Refs.~\cite{denton},~\cite{kuzmenko}, i.e. increases with $T$. It is possible if potentials $\lambda$,
$\lambda_{m\pm}$ comply with Eq.~(\ref{lambdaest}). At last, if a system is heated up to $T>\varepsilon_F$ and survives at such temperatures both description (canonical and equivalent grand canonical) lead to the Maxwell distribution, i.e. $n(\varepsilon)=f(\varepsilon)$, that can be provided if $Z_{EGCE}(N)/Z_{CE}(N)$ is a temperature independent constant. In this case, at $T>\varepsilon_F$ and $N\gg m$ it can be easy found that
\begin{equation}\label{lambdagam}                                                 %(30)
\lambda_{m+}\simeq\lambda_{m-}\simeq\lambda=-\gamma\ln\frac{T}{\varepsilon_F},
\end{equation}
$\gamma$ characterizes the increase of the average level density with energy
\begin{equation}\label{roav}                                                 %(31)
\rho_{av}(\varepsilon)=\gamma\frac{N}{\varepsilon_F}\left(\frac{\varepsilon}{\varepsilon_F}\right)^{\gamma -1},
\end{equation}
e.g. $\gamma=3;2;1$ for $3D$-, $2D$- and $1D$-oscillator and $\gamma =3/2;1;1/2$ for $3D$-, $2D$- and $1D$-rectangular potentials. Thus, the values of the chemical potentials, Eq.~(\ref{lambdagam}), are again in accordance with Eq.~(\ref{lambdaest}). Therefore we can infer that Eq.~(\ref{lambdaest}) is justified in a wide temperature range.

The estimations given by Eq.~(\ref{lambdaest}) lead to the well known fact(~\cite{landsberg}) that at low temperatures $n(\varepsilon\le\varepsilon_F)>f(\varepsilon\le\varepsilon_F)$ and $n(\varepsilon>\varepsilon_F)<f(\varepsilon>\varepsilon_F)$ i.e. the stepwise variation of the canonical occupation numbers ($n(\varepsilon)$) vs $\varepsilon$ near $\varepsilon_F$ is pronounced more distinctly as compared with the grand canonical ones ($f(\varepsilon)$).

The temperature derivative of $n(\varepsilon)$ i.e. $\partial n(\varepsilon)/\partial T=-\beta^2\partial n(\varepsilon)/\partial\beta$ is formally similar to $\partial f(\varepsilon)/\partial T$ if the latter is expanded in powers of $exp\left [\pm\beta(\varepsilon-\lambda)\right]$, Eqs.~(\ref{felt}) and (\ref{fgt}). The distinction consists again in the chemical potentials: $\lambda_{m\pm}$ in $n(\varepsilon)$ and the only value of $\lambda$ in $f(\varepsilon)$, Eq.~(\ref{dfdb}).
\begin{eqnarray}\label{dneltdb}
\frac{\partial n(\varepsilon\le\varepsilon_F)}{\partial T}=\nonumber\\
=\beta^2\sum_{m=1}^{M-N}(-)^{m+1}m\left(\varepsilon-\lambda_{m+}-\beta\frac{\partial\lambda_{m+}}{\partial\beta}\right)
\exp{\left [\beta m(\varepsilon-\lambda_{m+})\right]};
\end{eqnarray}
\begin{eqnarray}
\frac{\partial n(\varepsilon >\varepsilon_F)}{\partial T}=\nonumber\\
=\beta^2\sum_{m=1}^{N}(-)^{m+1}m\left(\varepsilon-\lambda_{m-}-\beta\frac{\partial\lambda_{m-}}{\partial\beta}\right)
\exp \left[-\beta m(\varepsilon-\lambda_{m+})\right];\label{dngtdb}
\end{eqnarray}
\begin{eqnarray}
m\left(\lambda_{m\pm}+\beta\frac{\partial\lambda_{m\pm}}{\partial\beta}\right)=
\mp\frac{\partial}{\partial\beta}\left [\ln Z(N\pm m)-\ln Z(N)\right ]=\nonumber\\
=\pm\left[ E(N\pm m)-E(N)\right ].\label{mlambet}
\end{eqnarray}
Eq.~(\ref{mlambet}) indicates that for high $N$ and $T$ (when $\partial E/\partial N$ makes sense)
\begin{eqnarray}
\lambda_{m\pm}+\beta\frac{\partial\lambda_{m\pm}}{\partial\beta}\longrightarrow\frac{\partial E}{\partial N}\nonumber
\end{eqnarray}
that is in accordance with Eqs.~(\ref{dEdN}) and (\ref{entropy}).

The relationship between the chemical potentials $\lambda_{m\pm}$ and $\lambda$ establishes the general tendency for differences $E_{EGCE}-E_{CE}$ and $S_{EGCE}-S_{CE}$ since function $R$ in Eq.~(\ref{ZEGCEpol}) owing to Eq.~(\ref{lambdaest}) is an increasing function of $T$ ($R\longrightarrow 1$ at $T\longrightarrow 0$; $R\longrightarrow M+1$ at $T\longrightarrow\infty$) or a decreasing function of $\beta=T^{-1}$. This gives rise to
\begin{eqnarray}                                        %(34)
E_{EGCE}=-\frac{\partial}{\partial\beta}\ln Z_{EGCE}=E_{CE}-\frac{1}{R}\frac{\partial R}{\partial\beta}>E_{CE}\\
-\frac{\partial R}{\partial\beta}=
\sum_{m=1}^N m\left [(\lambda+\beta\frac{\lambda}{\partial\beta})-(\lambda_{m-}+\beta\frac{\lambda_{m-}}{\partial\beta})\right ]
e^{\beta m(\lambda-\lambda_{m-})}+\nonumber\\
\sum_{m=1}^{M-N}m\left [(\lambda_{m+}+\beta\frac{\lambda_{m+}}{\partial\beta})-(\lambda+\beta\frac{\lambda}{\partial\beta})\right ]
e^{-\beta m(\lambda_{m+}-\lambda_{m})}>0\label{dRdb}\\
S_{EGCE}=S_{CE}+\ln R-\frac{\beta}{R}\frac{\partial R}{\partial\beta} > S_{CE}\nonumber
\end{eqnarray}
As at low $T$ only several first terms in Eq.~(\ref{dRdb}) are essential the positive definiteness of $(-\partial R/\partial\beta)$ results in
\begin{equation}\label{lambdaneq}
\lambda_{m+}+\beta\frac{\partial\lambda_{m+}}{\partial\beta}\ge\lambda+\beta\frac{\partial\lambda}{\partial\beta}
\ge\lambda_{m-}+\beta\frac{\partial\lambda_{m-}}{\partial\beta}
\end{equation}
at least for small values of $m$ and low $T$.

For calculations of the heat capacity $C$ and for qualitative explanation of temperature and particle number dependencies of $C$ at low $T$ it is expedient to introduce into consideration the spectral heat capacity distribution $\varphi(\varepsilon)$ which at $\varepsilon=\varepsilon_s$ gives the contribution of one electron on level $s$ to $C$
\begin{eqnarray}\label{Cfi}                                         %(34)
C=\sum_s d_s\varphi(\varepsilon_s)=\int\rho(\varepsilon)\varphi(\varepsilon)d\varepsilon\\
\varphi(\varepsilon)=(\varepsilon -\varepsilon_0)\frac{\partial n(\varepsilon)}{\partial T}\label{fi}            %(35)
\end{eqnarray}
An arbitrary constant $\varepsilon_0$ does not affect the values of $C$ as
\begin{eqnarray}
\int\rho(\varepsilon)\frac{\partial n(\varepsilon)}{\partial T}d\varepsilon=0\nonumber\\
\end{eqnarray}
For convenience' sake of comparison with $EGCE$ calculations below we keep $\varepsilon_0$ to be equal to ($\lambda+\beta\partial\lambda/\partial\beta$). An analogous spectral heat capacity for $EGCE$ can be introduced on the basis of  Eq.~(\ref{Cegce})
\begin{eqnarray}
\varphi_{eff}(\varepsilon)=(\varepsilon -\lambda -\beta\frac{\partial\lambda}{\partial\beta})\frac{\partial f(\varepsilon)}{\partial T}
=\beta^2(\varepsilon -\lambda -\beta\frac{\partial\lambda}{\partial\beta})^2f(\varepsilon)\left[1-f(\varepsilon)\right]\label{fieff}   %(36)
\end{eqnarray}
The main feature of both function $\varphi(\varepsilon)$ and $\varphi_{eff}(\varepsilon)$ important to interpreting $C$ at low $T$ consists in that these functions being depicted vs energy turn out to be two humped curves. Such function dependence on $\varepsilon$ can be easily exposed for $\varphi_{eff}(\varepsilon)$ that is given analytically, Eq.~(\ref{fieff}), and determined by only two parameters $\lambda$ and $\partial\lambda/\partial\beta$ (the latter at low $T$ is much smaller than $\lambda$).
In fact such form form of the curve for $\varphi_{eff}(\varepsilon)$ arises due to the crossings of the bell-wise curve $f(\varepsilon)\left[ 1-f(\varepsilon )\right]$ with the parabola $(\varepsilon -\lambda -\beta\partial\lambda/\partial\beta)^2$ and two maxima of $\varphi_{eff}(\varepsilon)$ at $\beta\partial\lambda/\partial\beta\ll\lambda$ are in points
\begin{eqnarray}\label{emax}                                         %(37)
E^{(eff)}_{max}(T)\simeq\lambda\pm 2.4T
\end{eqnarray}
i.e. the energy distance between the maxima is
\begin{eqnarray}\label{5T}                                         %(45)
\Delta E^{(eff)}_{max}(T)\simeq 4.8T.
\end{eqnarray}
Heating moves apart these maxima and thereby extends the energy space under the curve of $\varphi(\varepsilon)$.

The explicit form of $\varphi(\varepsilon)$ depends on the single-electron spectrum structure more essentially through the set of parameters $\lambda_{m\pm}$ and $\partial\lambda_{m\pm}/\partial\beta$. Nevertheless $\varphi(\varepsilon)$ is similar to $\varphi_{eff}(\varepsilon)$ on the whole though the curve for $\varphi(\varepsilon)$ passes under the curve for $\varphi_{eff}(\varepsilon)$ at a fixed $T$ on account of the relations between $\lambda$ and $\lambda_{m\pm}$, Eq.~(\ref{lambdaest}). The distinctions between $\varphi(\varepsilon)$ and $\varphi_{eff}(\varepsilon)$ are discussed in the following sections.

\section{Origin of the low temperature resonance in $C$}

The two humped shape of $\varphi(\varepsilon)$ and $\varphi_{eff}(\varepsilon)$ suggests an idea how the appearance of the low temperature maximum in the temperature dependence of $C$ (mentioned in Introduction) could be explained.  At rather low temperatures ($T<\delta_F$) both peaks of the spectral heat capacities can be so sharp that just under each of them there can be only one single-electron level (resonance levels). Sometimes a group of levels close together can take the role of the resonance level. In systems with even $N$ two such levels could be levels $F$ (completely filled at $T=0$) and $F+1$. In this case $\lambda +\beta\partial\lambda/\partial\beta\simeq (\varepsilon_F+\varepsilon_{F+1})/2$ and a local maximum in the temperature dependence of $C$ could appear if the positions of two maxima of $\varphi(\varepsilon)$ or $\varphi_{eff}(\varepsilon)$  coincide with the energies of these levels i.e. $F$ and $F+1$. The case when the maximum in $C$ can arise at the coincidence of only one peak of $\varphi(\varepsilon)$ with one level is given by such spectrum near $\varepsilon_F$ in which the level $F$ is half-filled at $T=0$ and two levels $F$ and $F+1$ are remote from other single-particle levels of the spectrum. Then $\lambda +\beta\partial\lambda/\partial\beta\simeq\varepsilon_F$ and an amplification of $C$ will be caused by the coincidence of the right peak of $\varphi(\varepsilon)$ with the position of level $F+1$.

The distance between the maxima in the spectral heat capacity depends on the temperature. For $\varphi_{eff}(\varepsilon)$, as shown in the previous section, Eq.~(\ref{5T}), it is $\simeq 4.8T$, for the canonical function $\varphi(\varepsilon)$ this distance is smaller, ($2T\div 4.8T$), that will be shown in this section.  Therefore only at a quite definite temperature that can be called the resonance temperature ($T_{res}$) the positions  either of two peaks of $\varphi(\varepsilon)$ or one of them and the minimum of $\varphi(\varepsilon)$ exactly correspond to energies of the electron levels. Hence it follows that $T_{res}$ is of order of some part ($0.2~\div~0.5$) of $\Delta\varepsilon$, the spacing between the resonance levels.

The condition $T=T_{res}$ is necessary but insufficient for the appearance of the local maximum in $C$  as it does not provide yet small values of the spectral heat capacity ($\varphi(\varepsilon)$ or $\varphi_{eff}(\varepsilon)$) at energies of single particle levels adjacent to the resonance ones. Indeed, if the peaks of $\varphi(\varepsilon)$ are not too narrow and other levels (with energies $\varepsilon_s$) divorced from resonance ones by spacings $\sim\Delta\varepsilon_{F+1,F}=\varepsilon_{F+1}-\varepsilon_{F}$ give noticeable values of $\varphi(\varepsilon_s)$ as compared with values of $\varphi(\varepsilon_F)$ and $\varphi(\varepsilon_{F+1})$ then rise in temperature above $T_{res}$ does not decrease the heat capacity and a local maximum in $C$ will be practical invisible. Though both the positions of peaks and their widths are determined by $T$ (increasing the temperature causes moving apart and broadening the peaks) the possibility of the local maximum appearance depends not only on $T$ but also on the level structure near $\varepsilon_F$.

Mesoscopic single-particle spectra are in general heterogeneous and local concentrations of levels (level bunchings) alternate with their rarefactions. However in some cases a spectrum near $\varepsilon_F$ can be considered practically uniform with equal spacings between levels (the ideal equal level spacing spectrum with level degeneracies $d=2$ is the spectrum of the $1D$-oscillator). For such spectrum with $d=2$ the canonical partition functions are given in Refs.~\cite{denton,kuzmenko}. In the case of the completely occupied Fermi level at $T=0$ these functions lead to the following equations for $\lambda_{m\pm}$, if only terms greater than $q^3$ are taken into account ($q=exp(-\beta\delta)$, $\delta$ is the level spacing):
\begin{eqnarray}
\lambda_{1+}=\varepsilon_{F}+\delta-T\ln a, \;\;\;\; \lambda_{1-}=\varepsilon_{F}+T\ln a,\nonumber\\
\label{lambda1}\\
\lambda_{2+}=\varepsilon_{F}+\delta,\hspace{2cm}\;\;\;\;\;\;\;\; \;\;\;\;\lambda_{2-}=\varepsilon_{F},\nonumber\\
\nonumber\\
a=2(1+q^2)(1+2q)^{-1},\nonumber\\
\lambda=\varepsilon_{F}+\delta /2=(\varepsilon_{F}+ \varepsilon_{F+1})/2.\label{lambda2}
\end{eqnarray}
It is easy to check that $\lambda_{m\pm}$, $\lambda_{m\pm}+\beta\partial\lambda_{m\pm}/\partial\beta$, ($m=1,2$) obtained in Eqs.~(\ref{lambda1}),~(\ref{lambda2}) for $T<\delta$ are satisfied Eqs.~(\ref{lambdaest}),~(\ref{lambdaneq}) even at $T=\delta$.

\begin{figure}[p]
\scalebox{0.5}
{\includegraphics{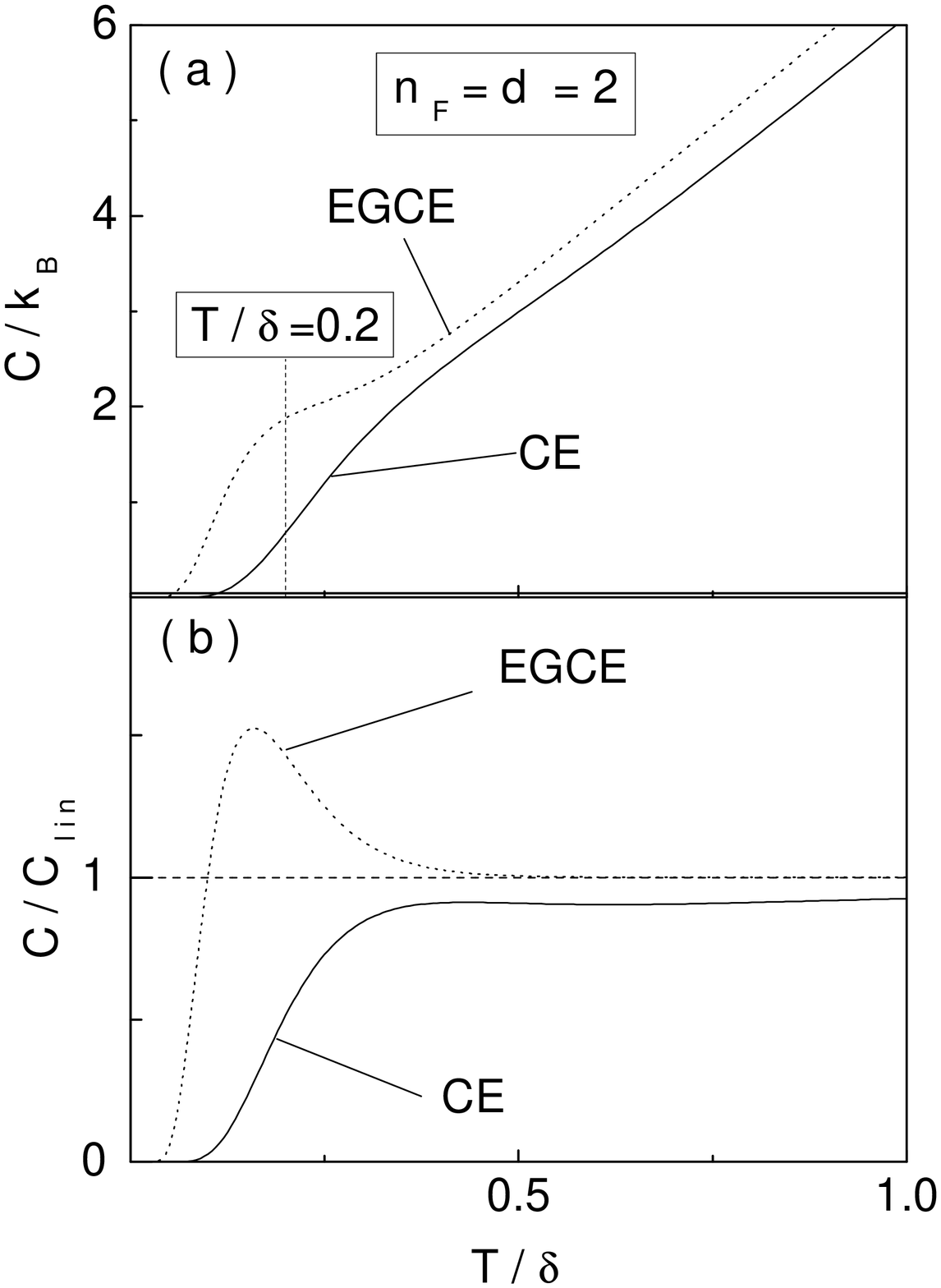}}
\caption{\label{DentCClin} The effective grand canonical and canonical heat capacity (a) and $C/C_{lin}$ (b) of $N$-even system vs $T$  in a model with the equal level spacing $\delta$, the level degeneracy of each level $d=2$. $n_F$ is the occupation number of the Fermi level at $T=0$.}
\end{figure}

\begin{figure}[p]
\scalebox{0.5}
{\includegraphics{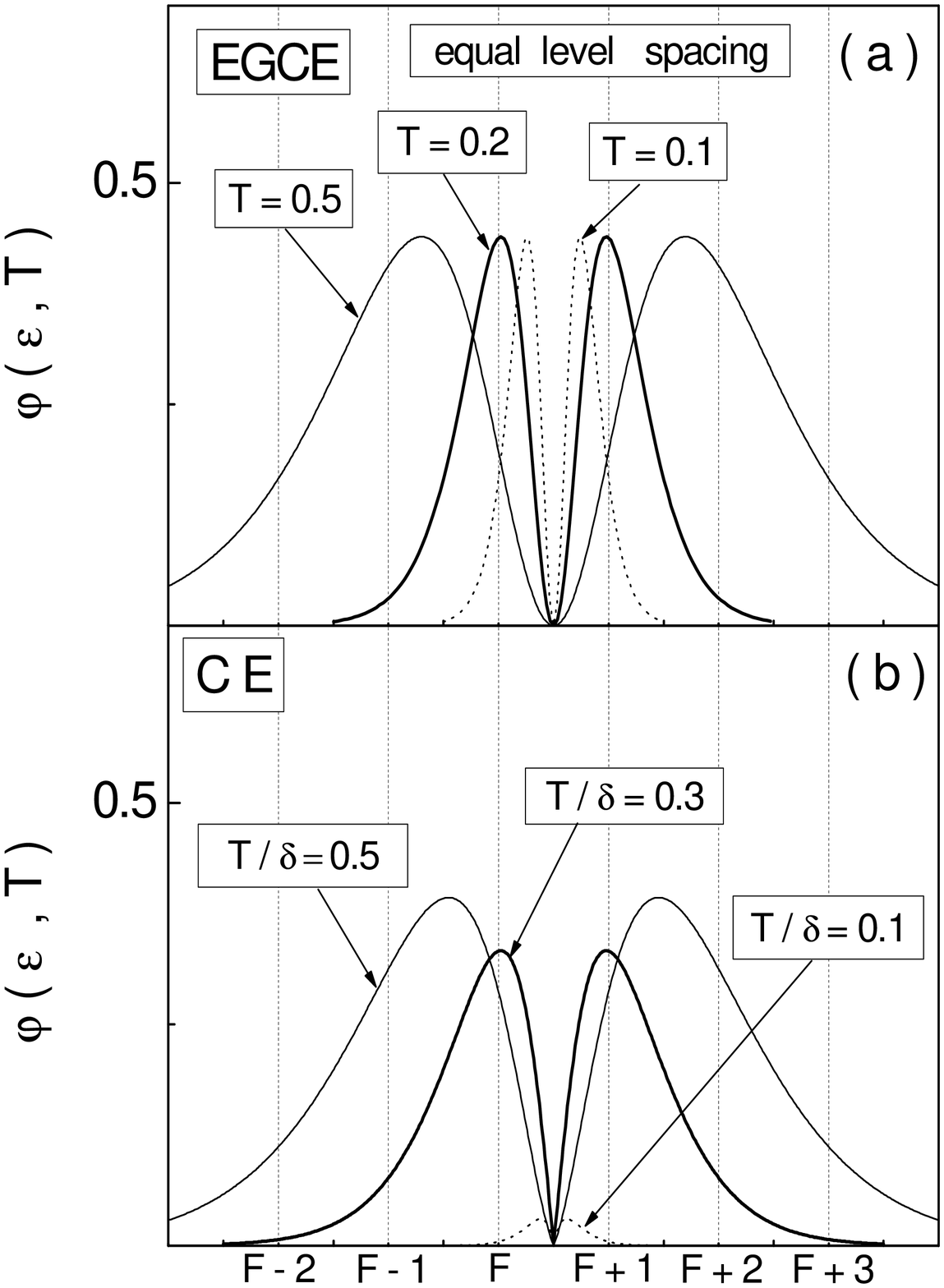}}
\caption{\label{FiEquald2} The spectral heat capacity in the $EGCE$ and $CE$  at different temperatures for the equal level spacing model. Vertical lines mark the level positions.}
\end{figure}

Since $\lambda_{1+}\simeq\lambda_{2+}$ ( the difference is small at low temperatures) and terms with $\lambda_{m+}$ ($m\geq 3$) are inessential at $T<\delta$ function $\varphi(\varepsilon)$ in the region $\varepsilon\leq\varepsilon_F$ can be approximated by $\varphi_{eff}(\varepsilon)$, Eq.~(\ref{fieff}), with $\lambda=\lambda_{1+}$. It means that the left peak of $\varphi (\varepsilon\leq\varepsilon_F)$ will be $\sim 2.4T$ distant from the point $\varepsilon_{F+1}\simeq\lambda_{1+}\simeq\varepsilon_{F}+\delta$ ($2.4T$ is the distance between the minimum and maximum of $\varphi_{eff}(\varepsilon)$ that gives $4.8T$ in Eq.~(\ref{5T})). Thus, the maximum of $\varphi (\varepsilon\leq\varepsilon_F)$ is closer to the point $\lambda$, Eq.~(\ref{lambda2}), than the peak of $\varphi_{eff}(\varepsilon)$ (the latter is $2.4T$ distant from $\lambda <\lambda_{1+}$). The analogous arguments point out that the right peak of $\varphi (\varepsilon >\varepsilon_F)$ is also closer to the point $\lambda$ than the right peak of $\varphi_{eff}(\varepsilon)$ at the same temperature. Thus, the peaks of $\varphi(\varepsilon)$ are roughly two times closer each other than the peaks of $\varphi_{eff}(\varepsilon)$. Hence the positions of the peaks of $\varphi(\varepsilon)$ will coincide with $\varepsilon_F$ and $\varepsilon_{F+1}=\varepsilon_{F}+\delta$ at $T_{res}\simeq 0.3\delta >T^{(eff)}_{res}$ as it displayed in Figs.~\ref{DentCClin},~\ref{FiEquald2}.
Fig.~\ref{FiEquald2} shows in addition that under the curve of $\varphi(\varepsilon)$ there are two more levels with noticeable (as compared with $\varphi(\varepsilon_F)=\varphi(\varepsilon_{F+1})$) values of $\varphi(\varepsilon_{F-1})$ and $\varphi(\varepsilon_{F+2})$. As mentioned above, it implies that the maximum in the canonical heat capacity will be practically absent  while in the $EGCE$ heat capacity it is well seen, Fig.~\ref{DentCClin}.
For odd $N$ systems with uniform spectra near $\varepsilon_F$ the temperature $T_{res}$ turns out to be even more than that for even ones as the chemical potential $\lambda$ in this case $\simeq\varepsilon_F$ while $\lambda_{m\pm}$ are practically the same as in Eqs.~(\ref{lambda1})~(\ref{lambda2}), i.e. the role of the resonance levels is performed by levels $F-1$ and $F+1$ that increases $T_{res}$, broadens the peaks of $\varphi(\varepsilon)$ and so the maxima in $C(CE)$ and $C(EGCE)$ are smoothed out.

In Fig.~\ref{DentCClin} and some following figures we give instead of $C$ vs $T$ values of $C/C_{lin}$ to more distinctly display appearing the local maximum in $C$ since in some cases such maximum looks more like a bend than a real maximum as it take place for $C(EGCE)$ vs $T$ in Fig.~\ref{DentCClin}a. $C_{lin}$ and $\rho_0(\varepsilon_F)$ are given by Eqs.~(\ref{Clin1}) and (\ref{roav}) respectively.

\begin{figure}[p]
\scalebox{0.5}
{\includegraphics{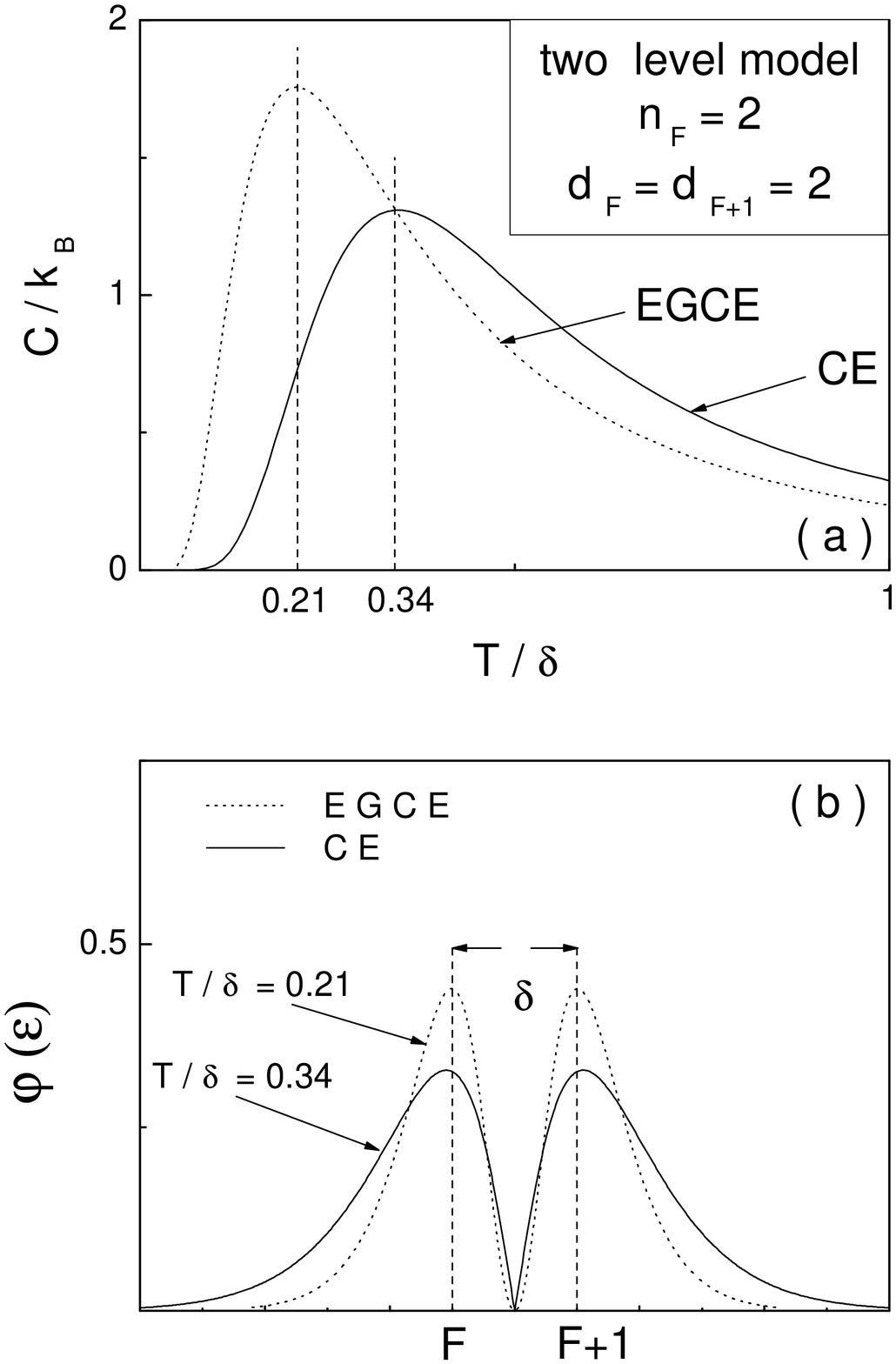}}
\caption{\label{CFi2levn2d2} Top panel: The heat capacity in the $EGCE$ and $CE$ vs $T$ in the two level model with $n_F=2$ and $d=2$. Vertical lines mark the resonance temperatures. Bottom panel: The spectral heat capacity in the $EGCE$ and $CE$  for resonance temperatures.}
\end{figure}

\begin{figure}[p]
\scalebox{0.5}
{\includegraphics{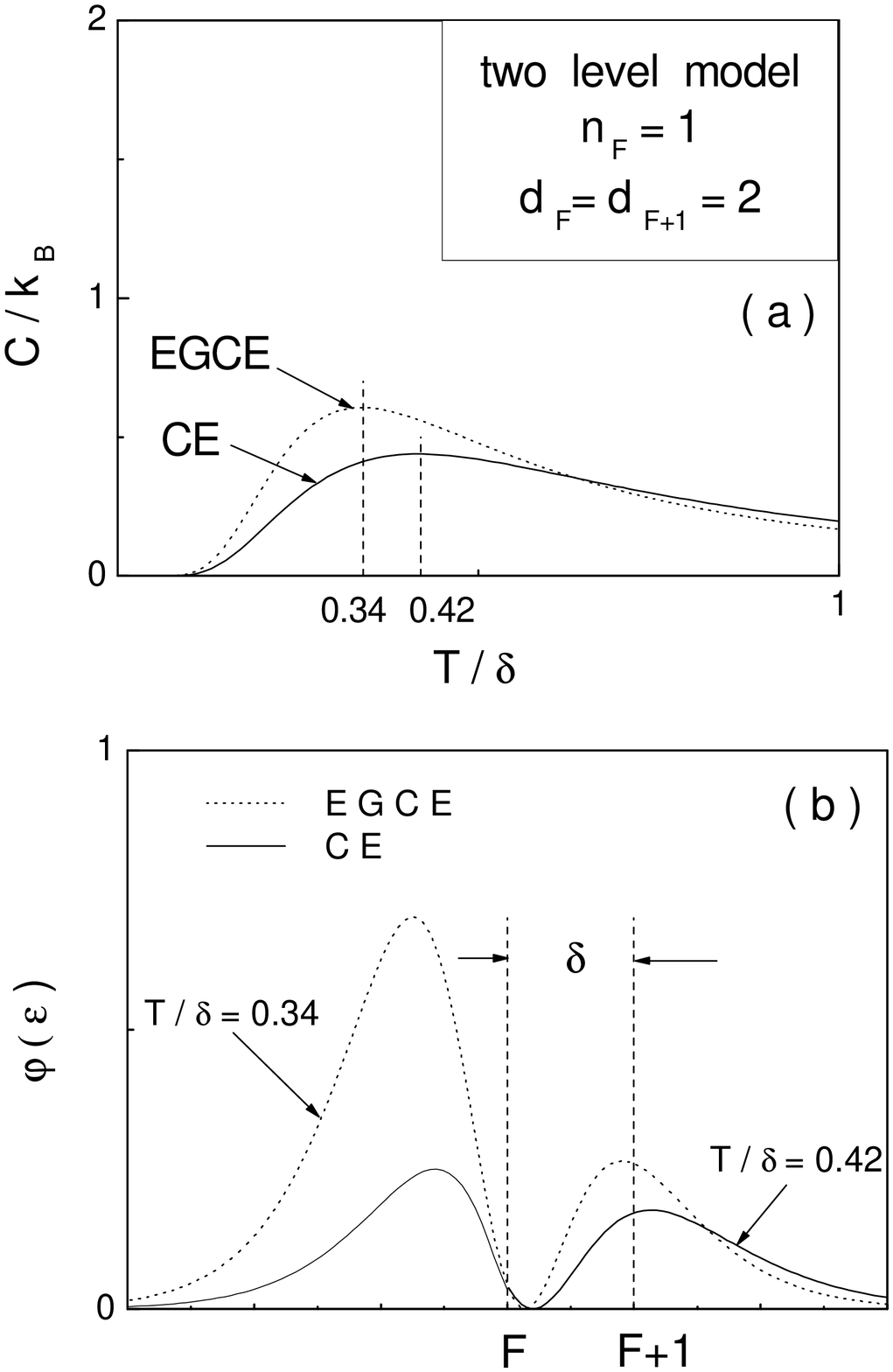}}
\caption{\label{CFi2levn1d2} Top panel: The heat capacity in the $EGCE$ and $CE$ vs $T$ in the two level model with $n_F=1$ and $d=2$.  Vertical lines mark the resonance temperatures. Bottom panel: The spectral heat capacity in the $EGCE$ and $CE$  for resonance temperatures.}
\end{figure}

Looking at Fig.~\ref{FiEquald2} one can assume that increasing a gap between levels $F$ and $F-1$ and simultaneously between $F+1$ and $F+2$ reduces contributions of levels $F-1$ and $F+2$ to $C$  and so the maximum in $C$ can arise. This is confirmed by calculations of $C$ for a system with two isolated levels ($F$ and $F+1$) where $F$ is completely occupied at $T=0$ (with $n_F=d_F=2$), Fig.~\ref{CFi2levn2d2}. In this case $T_{res}\simeq 0.3(\varepsilon_{F+1}-\varepsilon_{F})$.
For the same system with one electron on $F$ at $T=0$ ($n_F=1$) the maximum also exists but $T_{res}\simeq 0.4(\varepsilon_{F+1}-\varepsilon_{F})$ because now only the right peak of $\varphi(\varepsilon)$ coincides with level $F+1$ and the minimum of $\varphi(\varepsilon)$ practically falls on the point $\varepsilon_{F}$, Fig.~\ref{CFi2levn1d2}.

\begin{figure}
\scalebox{0.5}
{\includegraphics{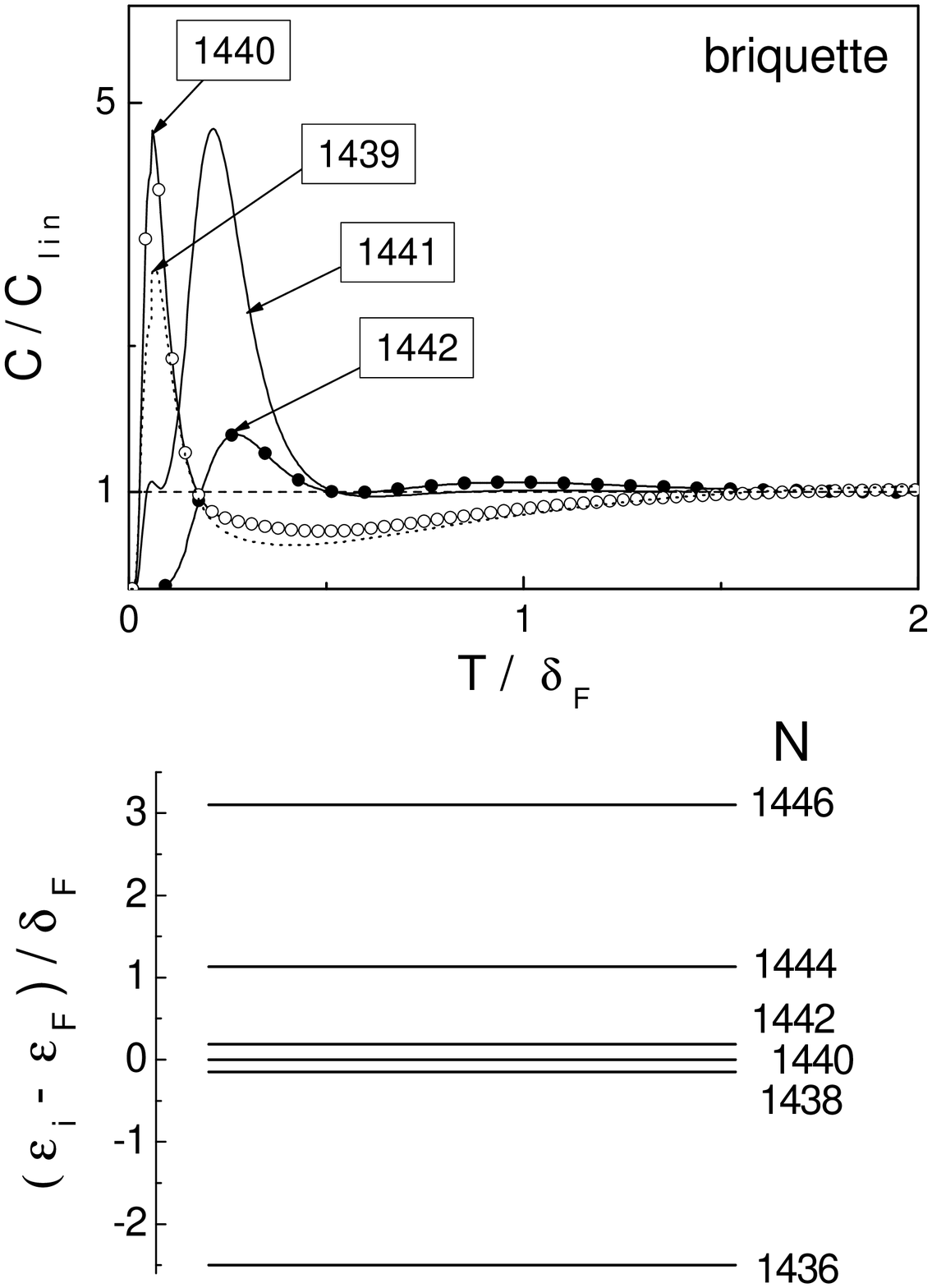}}
\caption{\label{Br14391442} Top panel: The heat capacity (in units of $C_{lin}$) of $N$-even and $N$-odd systems vs $T/\delta_F$ for a briquette. The lateral lengths  ($L_x:L_y:L_z=1:0.6e:\pi$) are chosen so to provide only the spin level degeneration, $\delta_F=4\varepsilon_F/3N$. Bottom panel: Fragment of the single-particle spectrum of the $N=1440$ briquette.}
\end{figure}

The example of a three level bunching generating local maxima in $C$ for both even and odd systems is given in  Fig.~\ref{Br14391442}. The single-electron spectrum producing the heat capacity variations and given in this figure belongs to a briquette-shaped cavity with hard walls and with such ratio of the lateral lengths that removes accidental degeneracies i.e. each level is only spin degenerated ($d=2$).

Thus, in asymmetric system having as a rule only spin degeneracies of levels the low temperature maximum in $C$ serve as evidence of a level concentration near $F$ and the resonance temperatures (at which the maxima is observed) give information concerning level spacing $\Delta\varepsilon$ ($\Delta\varepsilon$ can be equal to $\varepsilon_{F+1}-\varepsilon_{F}$ or $\varepsilon_{F+1}-\varepsilon_{F-1}$)
\begin{equation}\label{Trescan2}
T_{res}\sim (0.2\div 0.4) \Delta\varepsilon.
\end{equation}
Additional investigation of particle number variations of $C$ can help to ascertain detailed structure of the single-particle spectra.

In the single-particle spectra of symmetric $2D$-, and $3D$-systems the majority of states is high degenerated. Such extreme concentrations of levels affects the low temperature variation of $C$ via two factors both of which promote the appearance of the local maximum in $C$. The first is the extension, in the average, of spacings between high degenerated levels in comparison with asymmetric systems. The second is the approach of the canonical chemical potentials $\lambda_{m\pm}$, Eqs.~(\ref{expm}),(\ref{mlambda}), to the grand canonical value $\lambda$ with increasing degeneracy $d$ (i.e. increasing $N$) that decreases the difference between $\varphi(\varepsilon)$ and $\varphi_{eff}(\varepsilon)$ and makes more probable the appearance of the maxima in $C$ as the peaks in $\varphi_{eff}(\varepsilon)$ are rather narrow. At very high $d$ the canonical potentials become equal to $\lambda$ that results in identifying the canonical and effective grand canonical values of $\varphi(\varepsilon)$ and $C$.

As an illustration of these properties of symmetric systems we will consider a model example in which two solitary equally degenerated levels $F$ and $F+1$ are divided by the spacing $\delta$ and at $T=0$ the level $F$ is engaged $(n_F=d_F\gg 1$).

The polynomial method~\cite{kuzmenko} allows the ratio of the partition functions, Eqs.(\ref{expm}), to be written through finite sums in powers of $q$:
\begin{eqnarray}
q=\exp(-\beta\delta );\nonumber\\
\frac{Z(N\pm m)}{Z(N)}=R_m\exp \left \{ \mp\beta\left [(\varepsilon_F + \frac{\delta}{2})\pm\frac{\delta}{2}\right ] \right\} ;\label{Zratio}\\
R_m=\frac{\widetilde{Z_m}}{\widetilde{Z_0}};\;\;\; \widetilde{Z_m}=\sum_{k=0}^{d}
\left(
\begin{array}{c}
d  \\
m+k
\end{array} \right).
\left(
\begin{array}{c}
d  \\
k
\end{array} \right)
q^k;\nonumber\\
0\le m\le d.\nonumber
\end{eqnarray}

Hence, the canonical chemical potentials $\lambda_{m\pm}$ gain the form
\begin{equation}\label{lambdapm}
\lambda_{m\pm}=\lambda\pm\left [ 1-2(\beta\delta m)^{-1}\ln R_m\right ]\frac{\delta}{2},\;\;\;\; \lambda=\varepsilon_F + \frac{\delta}{2}.
\end{equation}
Since at $d\gg 1$ values of $R_m$ are proportional to $d^{m}$ the potentials $\lambda_{m\pm}$ explicitly depend on $\ln d$ i.e. as it is followed from Eq.(\ref{lambdapm}) the more $d$ the closer $\lambda_{m\pm}$ to $\lambda$. However the general tendency given by Eq.~(\ref{lambdaest}) is not broken at all variations of $d$, that we demonstrate for two cases: $d^2q\le 1$ and $d^2q\gg 1$.

If degeneracy $d$ is not too high
\begin{equation}\label{degen}
1\ll d<\exp(\beta\delta /2);\;\;\; \ln d<\beta\delta /2
\end{equation}
then neglecting terms $\sim d^{-1}$ and introducing a new parameter $p$
\begin{equation}\label{paramp}
p=d^2\exp(-\beta\delta /2)<1
\end{equation}
one can retain in the sums for $R_m$ only terms $\le p^2$ that simplifies series for it:
\begin{eqnarray}
R_m=d^m(m!r_m)^{-1}\label{Rmrm}\\
r_m=\left ( 1+p+\frac{p^2}{4}\right )\left ( 1+\frac{p}{m+1}+\frac{p^2}{2(m+1)(m+2)}\right )^{-1}.\label{rm}
\end{eqnarray}
Thereby Eq.(\ref{lambdaest}) is satisfied as
\begin{eqnarray}
\lambda_{m\pm}=\lambda\pm (a+b)\frac{\delta}{2};\label{lambdapm2}\\
a=1-2(\beta\delta)^{-1}\ln d >0;\nonumber\\
b=2(\beta\delta k)^{-1}\left [\ln (k!)+\ln r_k \right ],\nonumber
\end{eqnarray}
the positive definiteness of $a$ and $b$ is the consequence of Eqs.~(\ref{degen}) and ~(\ref{Rmrm}),~(\ref{rm}).

In the case of very high degeneracies
\begin{equation}\label{highd}
d^2\exp(-\beta\delta)\gg 1
\end{equation}
series for $\widetilde{Z_m}$ in Eq.~(\ref{Zratio}) can be reduced (again neglecting terms $\sim d^{-1}\ll 1$) to the Bessel function of purely imaginary argument $I_m(y)$, $y=2dq^{1/2}$:
\begin{equation}\label{Zm2}
\widetilde{Z_m}\simeq q^{-m/2}I_m(y)=\sum_{k=0}^{\infty}\left [k!(m+k)!\right ]^{-1}\left (\frac{y}{2}\right )^{2m}
\end{equation}
Now recollecting that the Bessel function asymptotic behavior $(y\gg 1)$ does not depend on $m$ one obtains by using Eq.~(\ref{Zratio})
\begin{equation}\label{Rm2}
R_m=\exp(\beta\delta m/2)
\end{equation}
that implies on account of Eq.(\ref{lambdapm}) the identity of $\lambda_{m\pm}$ and $\lambda$. Thus, in the case of very high degeneracies (at $N>>1$) the canonical and effective grand canonical descriptions give the identical values of $\varphi(\varepsilon)$ and $C$.

\begin{figure}[p]
\scalebox{0.5}
{\includegraphics{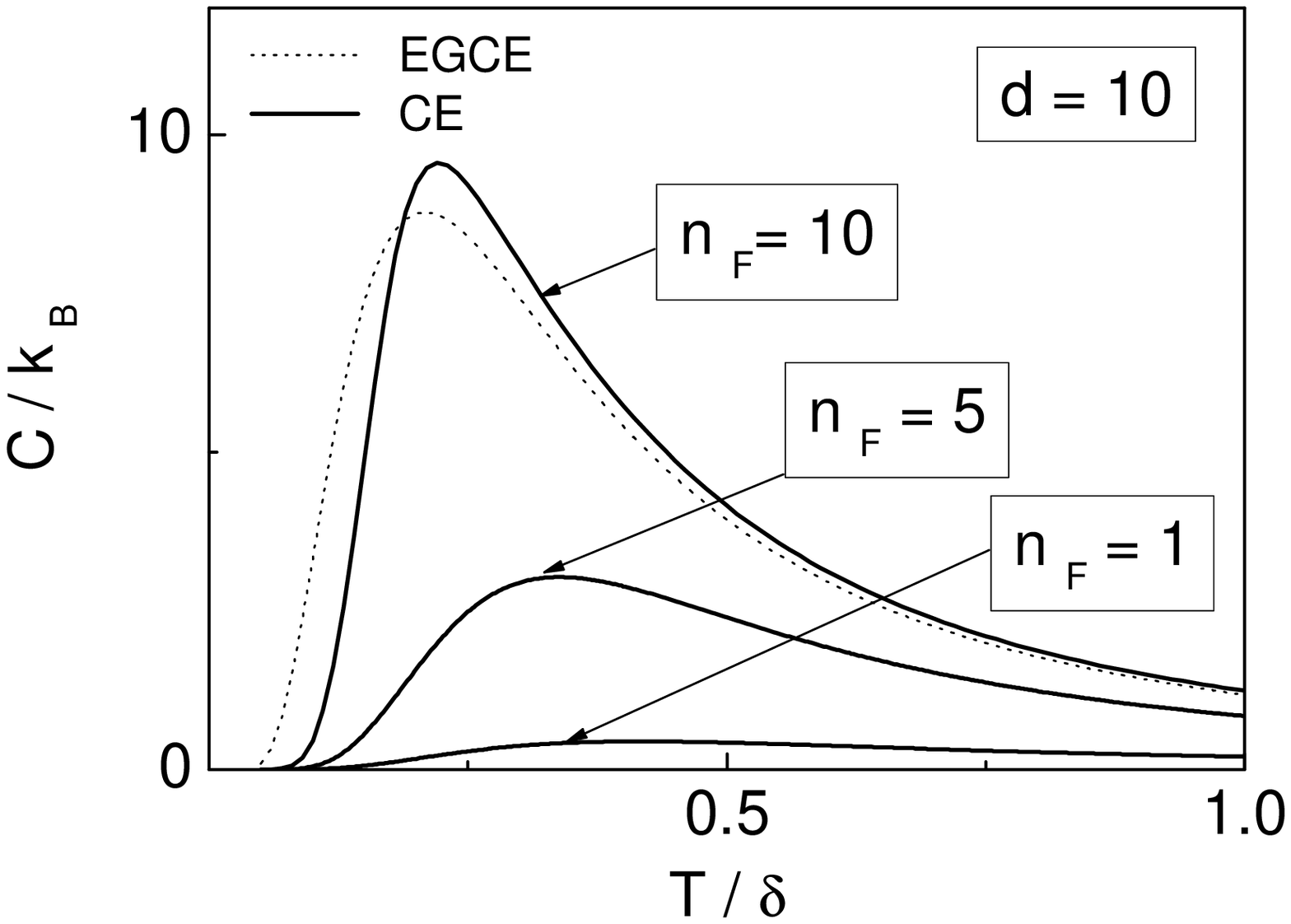}}
\caption{\label{C2levd10}  Impact of the Fermi level filling  on the low temperature resonance in the canonical heat capacity. Two level model, $d_F=d_{F+1}=10$.}
\end{figure}

Here it is worth mentioning one more detail distinguishing symmetric systems with $d\gg 1$ from asymmetric ones with $d=2$. In the latter $C_{CE}$ near $T_{res}$ is always less than $C_{EGCE}$ due to the relationship between the chemical potentials $\lambda_{m\pm}$ and $\lambda$, Eq.~(\ref{lambdaest}). However in systems with high level degeneracy approaching $\lambda_{m\pm}$ to $\lambda$ can cause a so small difference between these potentials that the pre-exponent factors in $C$ can compensate this difference and make $C_{CE}>C_{EGCE}$ in a narrow temperature range around $T_{res}^{(can)}$. E.g. in low temperature approximation the just considered two level model with $d^2\exp(-\beta\delta)<1$ and $d=n_F\gg 1$ gives:
\begin{equation}\label{Ccanlow}
C_{CE}\simeq (\beta\delta )^2d^2\exp(-\beta\delta); \;\;\;C_{EGCE}\simeq (\beta\delta )^2(d/2)\exp(-\beta\delta).
\end{equation}
For $d=n_F=10$, e.g., it means that $C_{CE}>C_{EGCE}$ in the vicinity of $T_{res}~\simeq~0.2~\delta$ as it is shown in Fig.~\ref{C2levd10}.

\begin{figure}[p]
\scalebox{0.5}
{\includegraphics{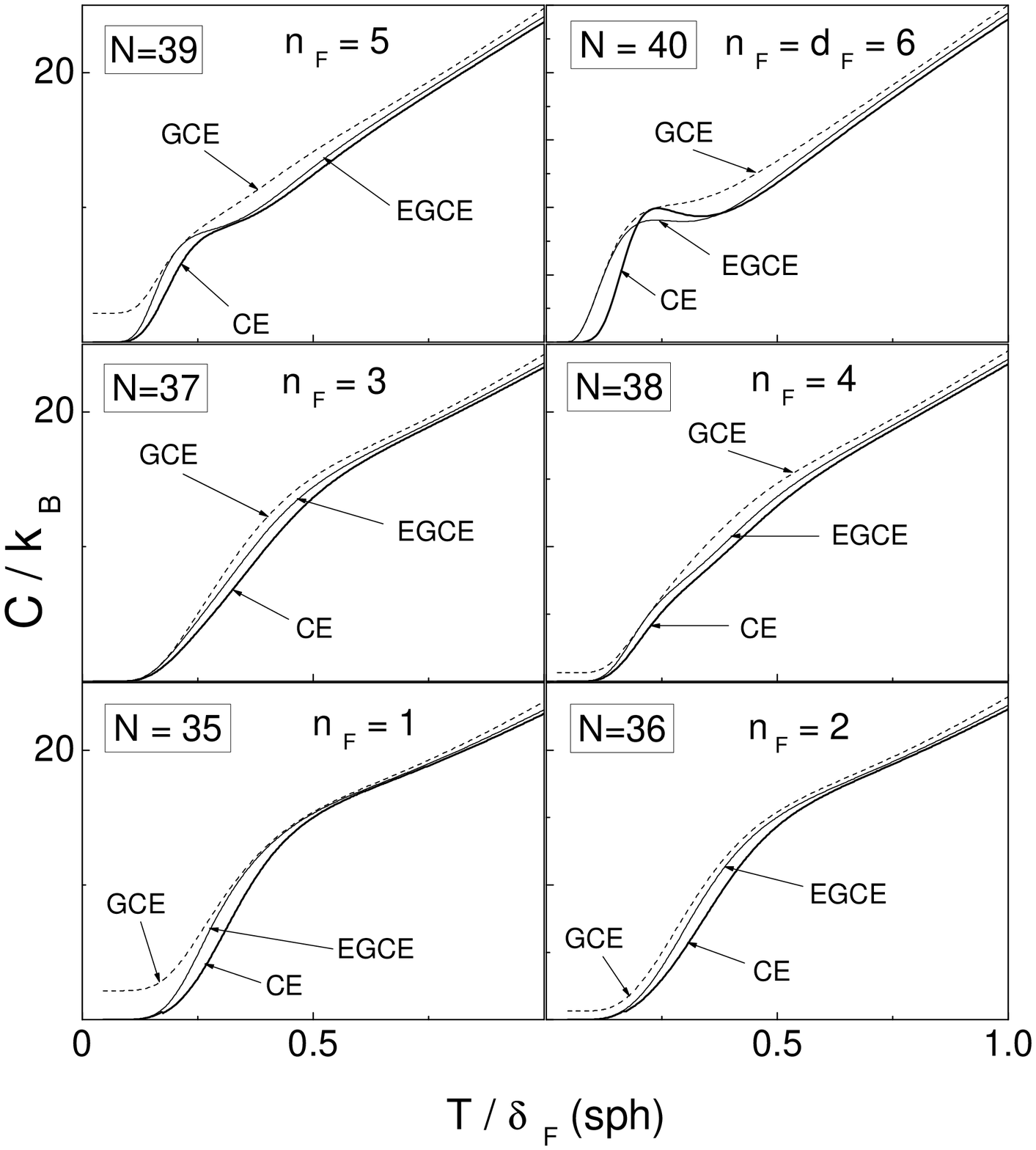}}
\caption{\label{CClin3540} The same as in Fig.~\ref{C2levd10} but for the $N=40$ shell in the spherical cavity, $n_F$ is the particle number at $T=0$ on the Fermi level with degeneracy $d_F=6$. For a sphere with $N=40$ the mean level spacing $\delta_F(sph)=\varepsilon_F/N^{1/2}\approx \varepsilon_{F+1}-\varepsilon_F$. Results of the canonical  ($CE$), equivalent grand canonical ($EGCE$) and grand canonical ($GCE$ in which $\lambda$ is temperature dependent but $\beta\partial\lambda/\partial\beta$ is omitted) methods are displayed. It is obvious that the $GCE$ method gives correct values of $C$ at $T\longrightarrow 0$ only at $n_F=d_F$ and $n_F=d_F/2$. In both other methods  $C\longrightarrow 0$ at $T\longrightarrow 0$.}
\end{figure}

\begin{figure}[p]
\scalebox{0.6}
{\includegraphics{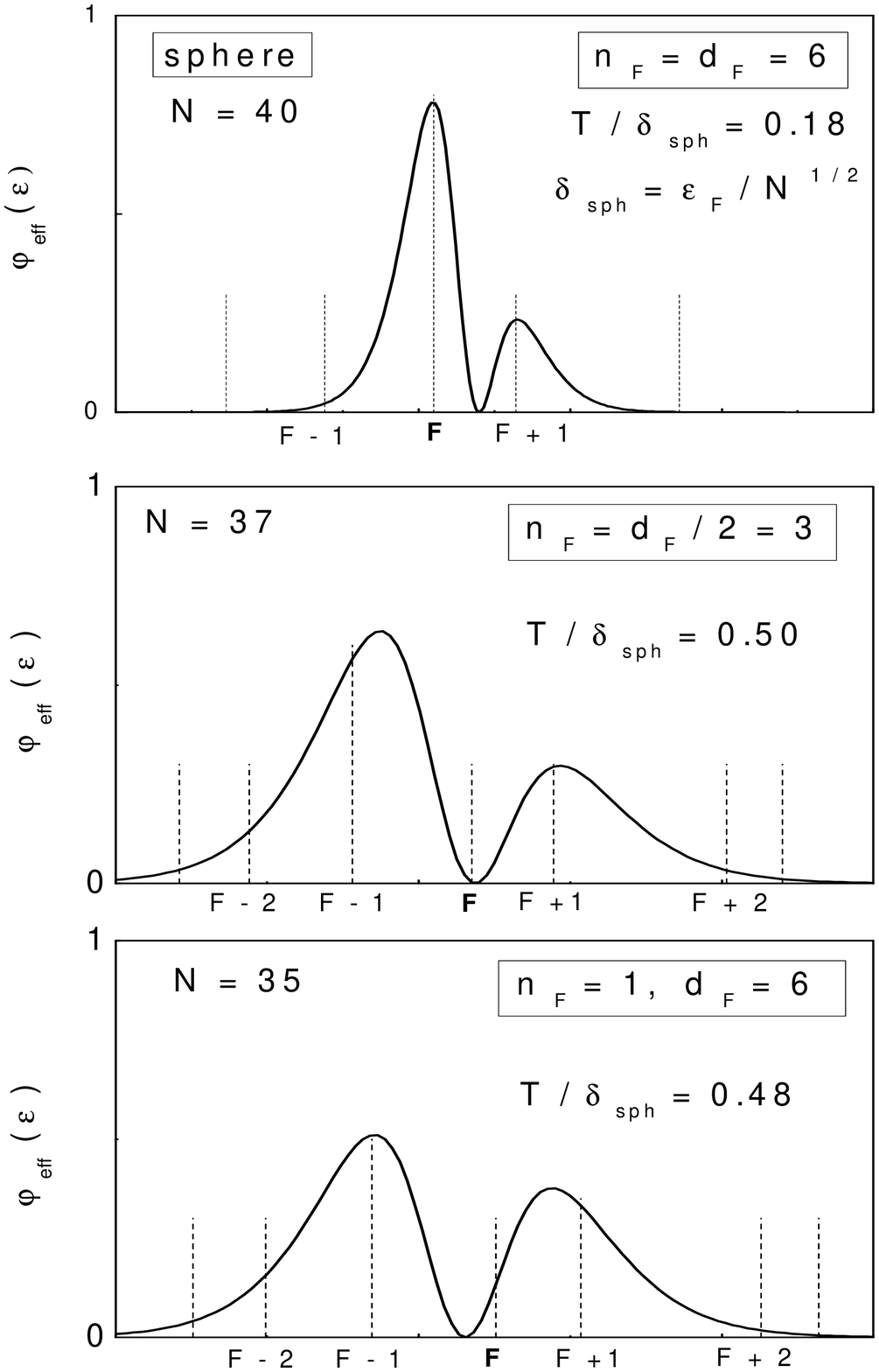}}
\caption{\label{FiGCE353740} The spectral distribution of the $EGCE$ heat capacity $\varphi_{eff}(\varepsilon)$ calculated for $n_F=1;3$ and $6$ (see Fig.~\ref{CClin3540}).}
\end{figure}

\begin{figure}[p]
\scalebox{0.5}
{\includegraphics{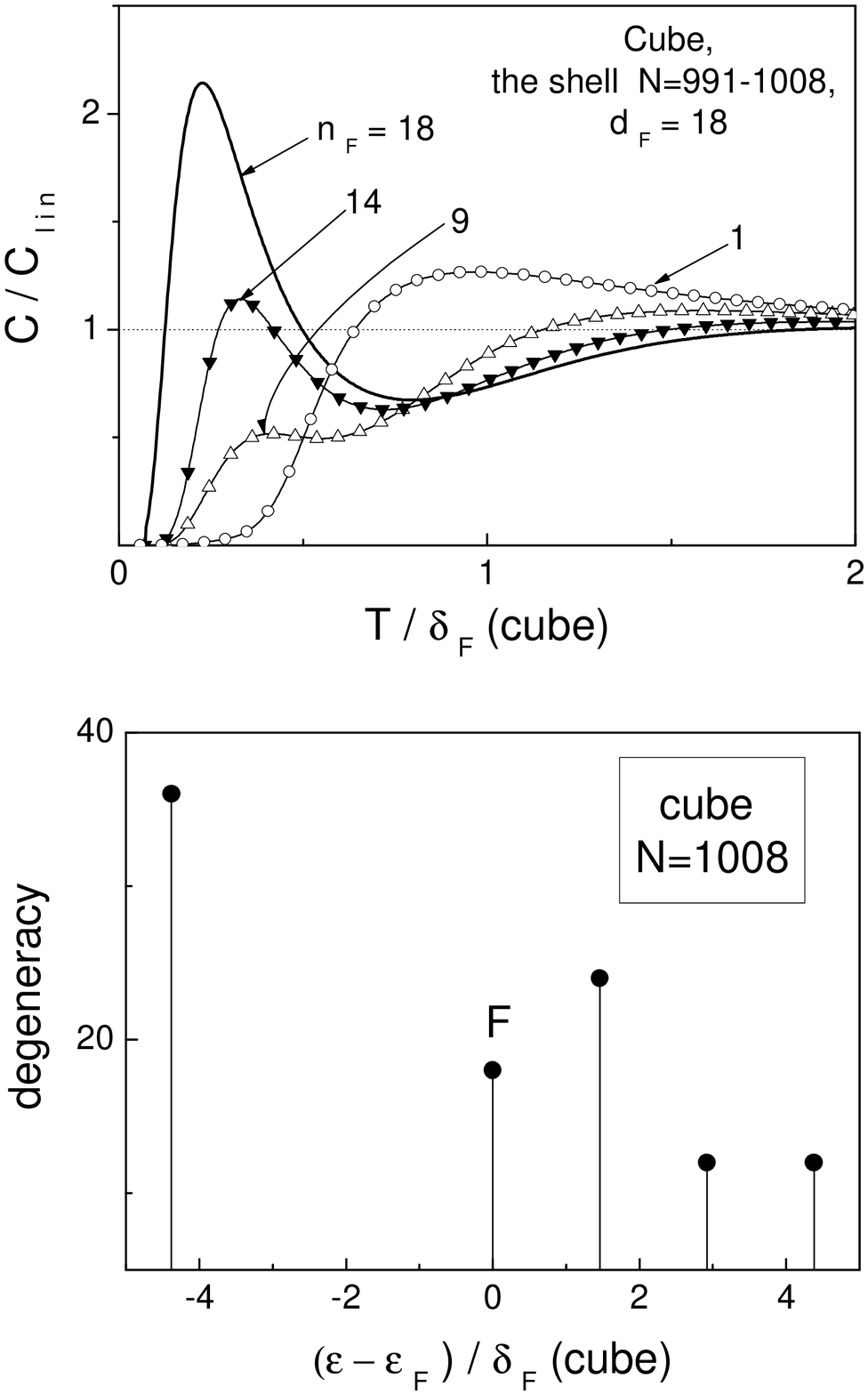}}
\caption{\label{Cube1008} Impact of the Fermi level filling  on the canonical heat capacity (in units of $C_{lin}$) for the shell of a cube with $N=991\div 1008$. $\delta_F(cube)=20\varepsilon_F/3N$ is adopted as the mean level spacing near the Fermi shell. For a cube with $N=1000$ $\delta_F (cube)\approx \varepsilon_{F+1}-\varepsilon_F$. $C_{EGCE}$ and $C_{CE}$ practically coincide since $d_F\gg 1$.}
\end{figure}

\begin{figure}[p]
\scalebox{0.5}
{\includegraphics{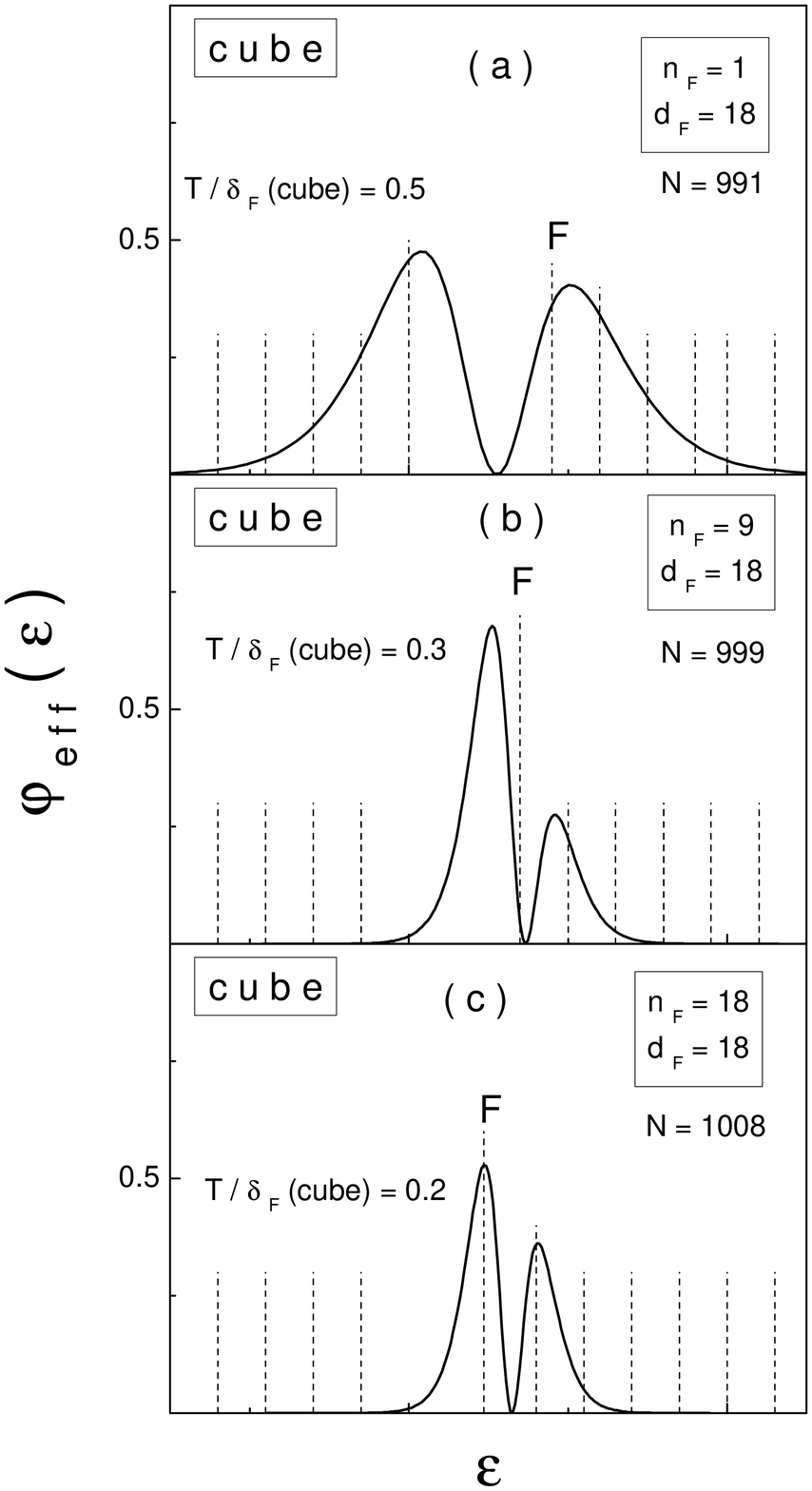}}
\caption{\label{FiEGCE1008} The spectral distribution of the effective grand canonical heat capacity $\varphi_{eff}(\varepsilon)$ calculated for $n_F=1;9$ and $18$ (see Fig.~\ref{Cube1008}).}
\end{figure}

\begin{figure}
\scalebox{0.6}
{\includegraphics{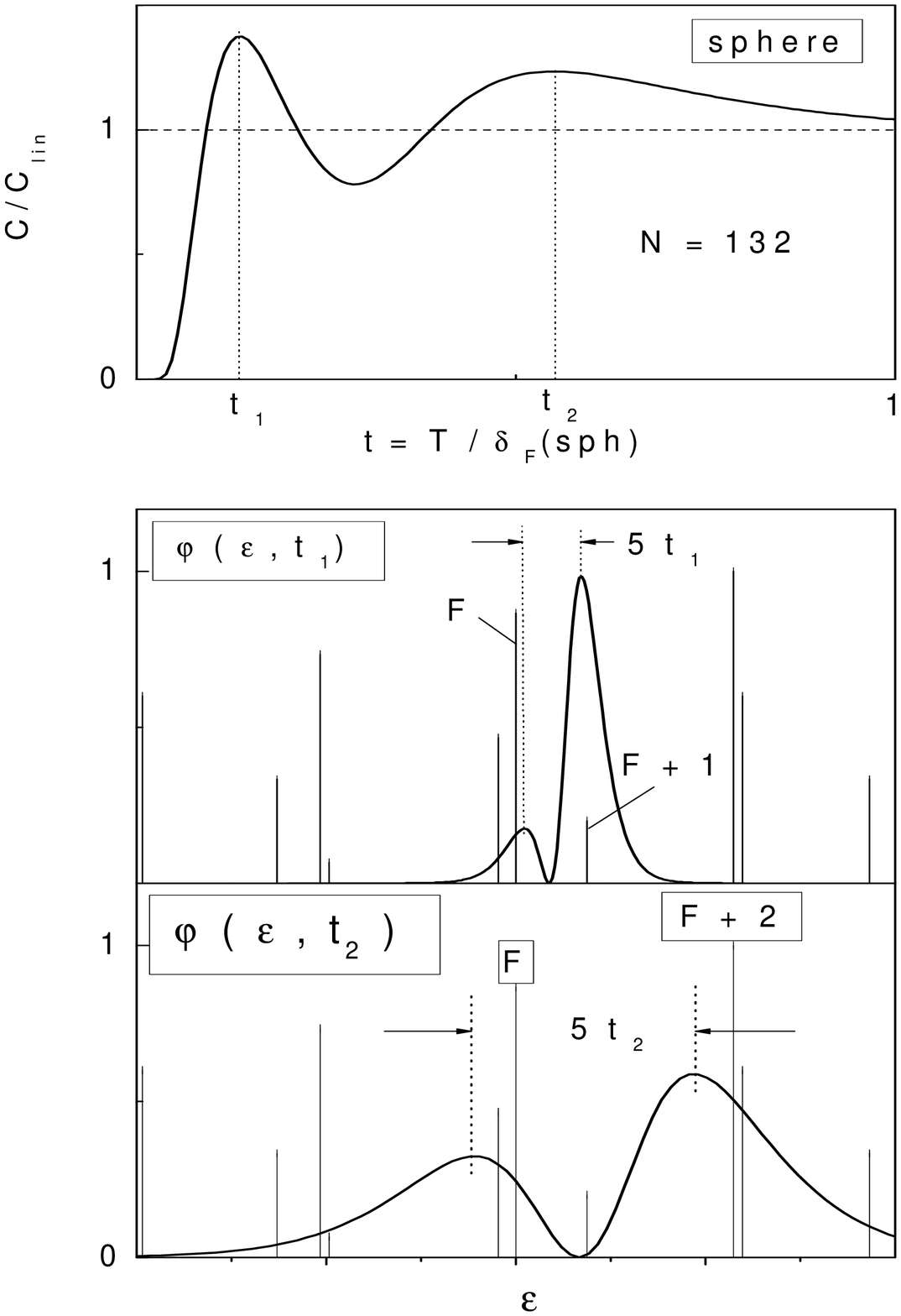}}
\caption{\label{2maxIntFunSph} Top panel: The heat capacity (in units of $C_{lin}$) with two resonances for $132$ fermions in spherical cavity. $t_1$ and $t_2$ are the positions of maxima. Bottom panel: The spectral heat capacity $\varphi_{eff}(\varepsilon ,t)$ at these temperatures. The first maximum is practically at $t_1=(\varepsilon_{F+1}-\varepsilon_F)/5\delta_F(sph)$ but $t_2$ corresponds to the presence of two groups of levels in the vicinity of the maxima in $\varphi_{eff}(\varepsilon ,t)$. Vertical lines mark the single-particle levels, their heights are proportional to the level degeneracies. $\delta_F(sph)=\varepsilon_F/N^{1/2}$ is the mean level spacing between spherical shells.}
\end{figure}

\begin{figure}
\scalebox{0.7}
{\includegraphics{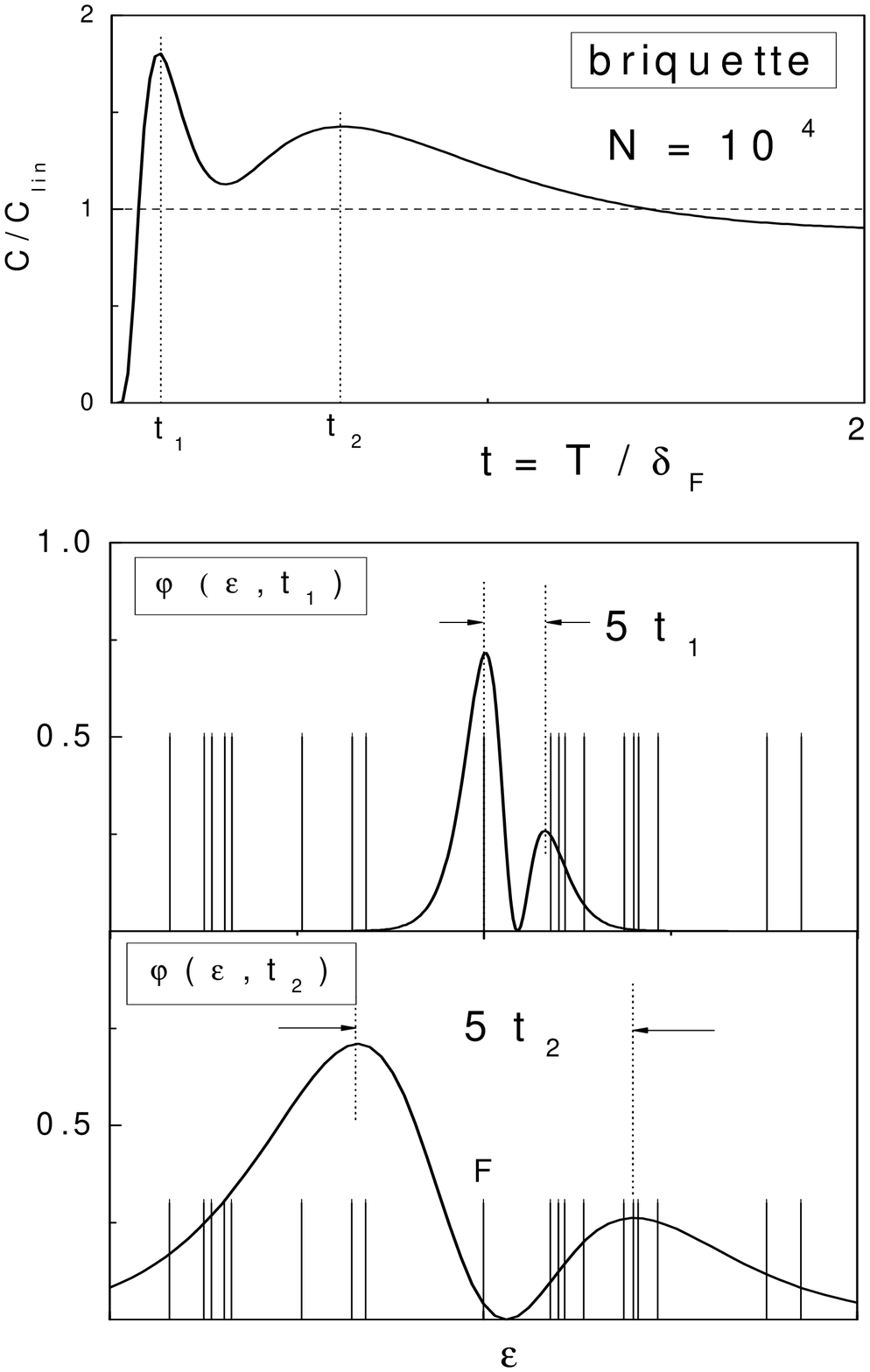}}
\caption{\label{2maxIntFunBr} The same as in Fig.~\ref{2maxIntFunSph} but for the $N=10^4$ briquette. $\delta_F=4\varepsilon_F/3N$}
\end{figure}

The local maximum appearance in $C$ in symmetric systems strongly depends on that to what extent the Fermi level is occupied at $T=0$, Fig.~\ref{C2levd10}. Since in such systems the spectral heat capacities $\varphi(\varepsilon)$ and $\varphi_{eff}(\varepsilon)$ are rather close, therefore, to simplify the problem we will address the later as it is characterized by only two parameters ($\lambda$ and $\beta\partial\lambda/\partial\beta$). If $T\sim T_{res}\simeq 0.2\delta_F$ the two level ($F$ and $F+1$) model is a reasonable approximation practically at any level distribution. In this model at $n_F=d_F$ the chemical potential $\lambda$ is between $\varepsilon_F$ and $\varepsilon_{F+1}$: if $d_F=d_{F+1}$ the value of $\lambda$ is just in the middle of this energy interval ($\lambda=(\varepsilon_F+\varepsilon_{F+1})/2$) otherwise it is closer to that level the degeneracy of which is lower. Decreasing $n_F$ up to $d_F/2$ makes the chemical potential and $\lambda +\beta\partial\lambda/\partial\beta$ approach to $\varepsilon_F$ i.e. in the regime $d_F\leq n_F\leq d_F/2$ decreasing $n_F$ gradually shifts the minimum of $\varphi_{eff}(\varepsilon)$ to $\varepsilon_F$. Subsequent decreasing $n_F<d_f/2$ leads to values of $\lambda$ appreciably lower than $\varepsilon_F$ but $\lambda +\beta\partial\lambda/\partial\beta$  holds its position near $\varepsilon_F$. The difference between these quantities (i.e. $\beta\partial\lambda/\partial\beta$) violates the left-right symmetry of $\varphi_{eff}(\varepsilon)$: its left peak becomes higher than the right one and the more is  $\beta\partial\lambda/\partial\beta$ the more distinct is the asymmetry of $\varphi_{eff}(\varepsilon)$. It implies that at $n_F\leq d_F/2$ level $F$ does not take part in forming $C$. If levels $F$ and $F+1$ are really far from other levels the maximum in $C$ cannot practically arise as the right peak of $\varphi_{eff}(\varepsilon)$ disposed over $\varepsilon_{F+1}$ is small and does not give a marked amplification of $C$. Such a case is displayed in Fig.~\ref{C2levd10}. This tendency reveals itself also in Fig.~\ref{CClin3540}. Interpretation of data in this figure can be carried out by using Fig.~\ref{FiGCE353740} where values of $\varphi_{eff}(\varepsilon)$ are given  for different $n_F$ in Fig.~\ref{CClin3540}. Fig.~\ref{CClin3540} shows also that the $GCE$ approach ($\lambda$ is temperature dependent but $\beta\partial\lambda/\partial\beta$ is omitted) does not give correct results for all $n_F$ at $T\longrightarrow 0$. Figs.~\ref{Cube1008},~\ref{FiEGCE1008} and Figs.~\ref{2maxIntFunSph},~\ref{2maxIntFunBr} show that for realistic spectra of spherical and cubic shells the forming of the local maxima involves not only $F$ and $F+1$ levels but also $F-1$ level.

In some cases heating mesoscopic bodies above $T_{res}$ can result in a local minimum in $C$ if the preceding local maximum was created by a single electron level group compactly disposed near $\varepsilon_{F}$ and far enough from other levels. Rise in temperature shifts therefore the peaks of $\varphi(\varepsilon)$ in energy regions free from single-electron levels that causes a decrease of $C$. Further heating can produce one more maximum in $C$ if one or several levels appear again under the peaks of $\varphi(\varepsilon)$ as it shown in Figs.~\ref{2maxIntFunSph},~\ref{2maxIntFunBr}. These figures indicate that the second resonance is formed by not one pair of levels, as it takes place for the first resonance ($5T_{res}\simeq\varepsilon_{F+1}-\varepsilon_F$) but in this case under the maxima of $\varphi(\varepsilon)$ there are at least two level bunches.

\section{Shape resonances in the fermion heat capacity}

\begin{figure}[p]
\scalebox{0.5}
{\includegraphics{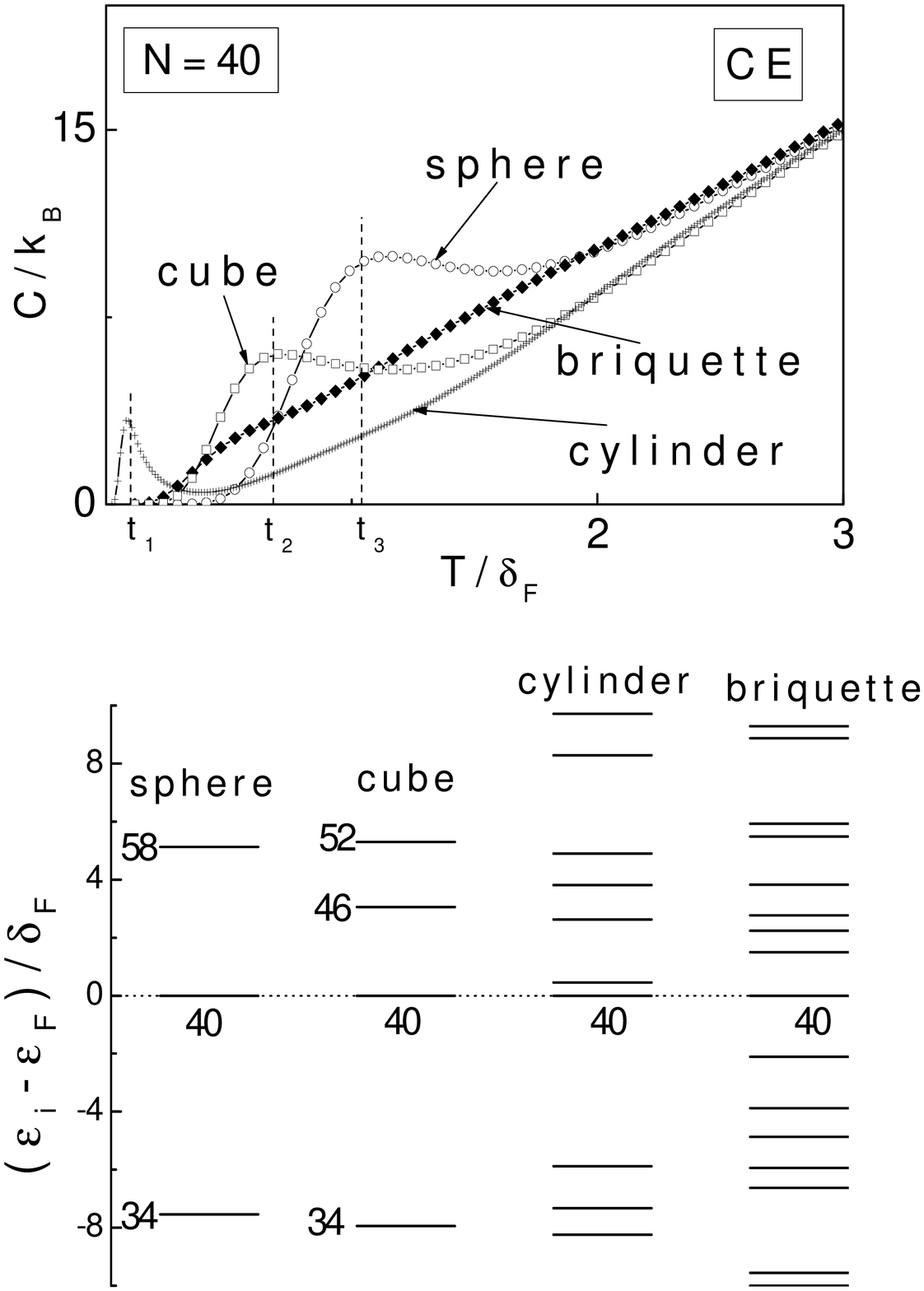}}
\caption{\label{LayCNa40cylsphbr}  Top panel: The heat capacities of  $N=40$ systems in $3D$-cavities of different shapes (briquette, cube, cylinder and sphere ). The dashed vertical lines mark values of  $t=T/\delta_F=0.2(\varepsilon_{F+1} - \varepsilon_F)/\delta_F$, Eq.~(\ref{5T}). These temperatures practically coincide with the positions of the corresponding maxima in $C$. In this figure $\delta_F=4\varepsilon_F/3N$. Bottom panel: Fragments of the single-particle level schemes of  sphere,  cube, cylinder (the diameter is equaled to the height) and briquette   ($L_x:L_y:L_z=1:1.1:1.2$).}
\end{figure}

\begin{figure}[p]
\scalebox{0.7}
{\includegraphics{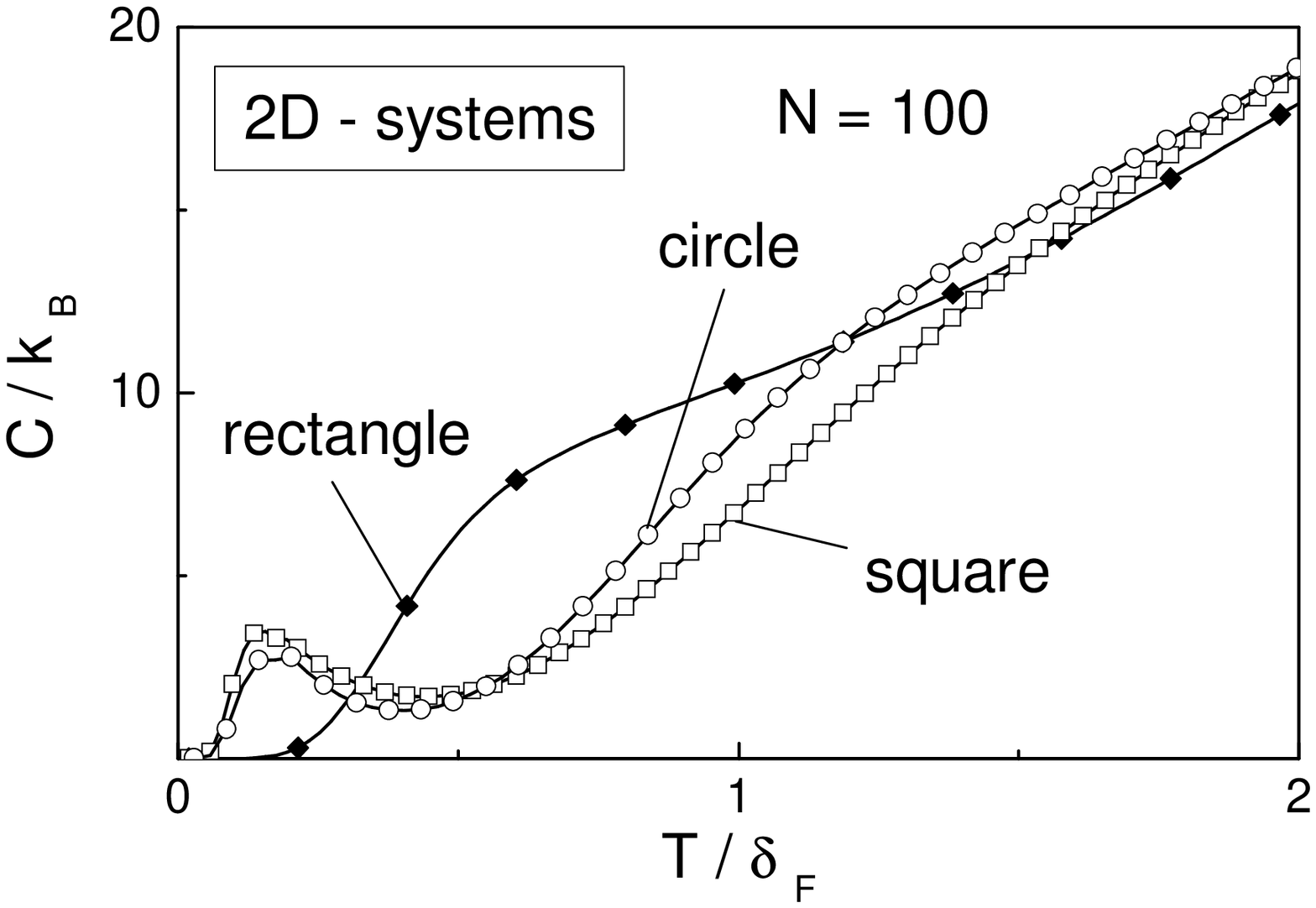}}
\caption{\label{CNa100circsqrect}  The heat capacities of  $N=100$ systems in $2D$-cavities of different shapes (circle, square, rectangle with $L_x:L_y=1:0.318$). $\delta_F=2\varepsilon_F/N$.}
\end{figure}

The canonical heat capacity temperature variations at $T<\delta$ described in the previous section in the framework of $NPM$ can serve as an indicator of the level distribution near $\varepsilon_F$: monotonous increase of $C_{CE}$ is associated with a uniform spectrum with $d=2$, extrema of $C$ give evidence concerning either local level concentration or high degeneracies of single-particle levels that is an inherent property of symmetric systems. The values of $C$ are also affected by the particle number as the complete occupation of the Fermi level at $T=0$ ($n_F=d_F$) in symmetric systems results in essential excess of $C$ over the Sommerfeld value $C_{lin}\sim N$. Thus, due to its dependence on the shape of the system and nonlinearity in the particle number the mesoscopic heat capacity at low temperatures paradoxically differs from $C$ of the same bodies at $T\gg\delta$ where the specific heat linearly increases with $T$ irrespective of the shape of the body. Examples of such shape dependence of $C$ is given in Fig.~\ref{LayCNa40cylsphbr} for $3D$-systems ($N=40$)  and in Fig.~\ref{CNa100circsqrect} for $2D$-systems ($N=100$)  which show the temperature variations of $C$ for $N$-particle systems of different but topologically equivalent shapes.


Another example of the heat capacity shape dependence is deformation oscillations in $C$. Before describing these oscillations from the point of view of $NPM$ we will indicate that these and other low temperature oscillations in $C$ are caused generally only by the quantization of the mesoscopic system excitation energies.

To develop this conception consider a mesoscopic many particle system subjected to an external impact that does not affect the volume of the system and thereby the particle number and that is characterized by a continuously varying parameter $\alpha$. The latter can be an external field strength, deformation parameter or something else. The hamiltonian or boundary conditions are supposed to be dependent on $\alpha$ in such a way that at some values of $\alpha$, $\alpha=\alpha_0$, one or several excited states of the system can merge with the ground state (level crossing) or, in other words, deviations of $\alpha$ from $\alpha_0$ can give rise to a splitting of the ground state into two or several components (sublevels). The energy of the splitting $\Delta E_{sp}(\alpha )$ at small $|\alpha -\alpha_0|$ has to be much smaller than $\Delta E$, the energy that separates the ground state at $\alpha=\alpha_0$ from other highly excited states, $\Delta E_{sp}(\alpha )\ll\Delta E$. Here the nature of the excitation spectrum i.e. what interactions between particles form it plays no role. The only property of the excited states which needs is their smooth variation determined by parameter $\alpha$. Then, increasing $\alpha$ to point $\alpha_0$ and further can cause at low temperatures, $T<\Delta E_{sp}(\alpha )$, the appearance of extrema in $C$. Indeed, at $T\ll \Delta E$ and $\alpha=\alpha_0$ the ground state is divided by $\Delta E$ from other states. Therefore the heat capacity is practically equal to zero ($\sim\exp [-\Delta E/T]$. Deviations of $\alpha$ from $\alpha_0$ create a doublet or multiplet of sublevels with energies $E_0$ ( a new ground state), $E_1,\ldots$ $E_i,\ldots$ $E_k$,  $E_i-E_0\leq\Delta E_{sp}(\alpha)=E_k-E_0$ ($i\leq k$) and the temperature transitions between these sublevels become possible. However two tendencies compete now: on the one hand increasing $|\alpha -\alpha_0|\sim E_i-E_0$ intensifies the contribution of state $i$ to $C\sim (E_i-E_0)^2$, but on the other hand the temperature factor $\sim\exp [-(E_i-E_0)/T]$ falls with increasing $(E_i-E_0)$. Consequently, at a fixed small temperature $(T<\Delta E_{sp}(\alpha)$) this competition can result in the appearance of two maxima in $C$ at values of $\alpha$ to the left and right of the point $\alpha_0$ and the mentioned above minimum of $C$ (in point $\alpha_0$) lying between these maxima. Thus, in a wide range of parameter $\alpha$ variations values of $C$ will oscillate: a pair of maxima will appear every time when $\alpha$ passes through each consequent point in which the level crossing occurs.
\begin{figure}[p]
\scalebox{0.5}
{\includegraphics{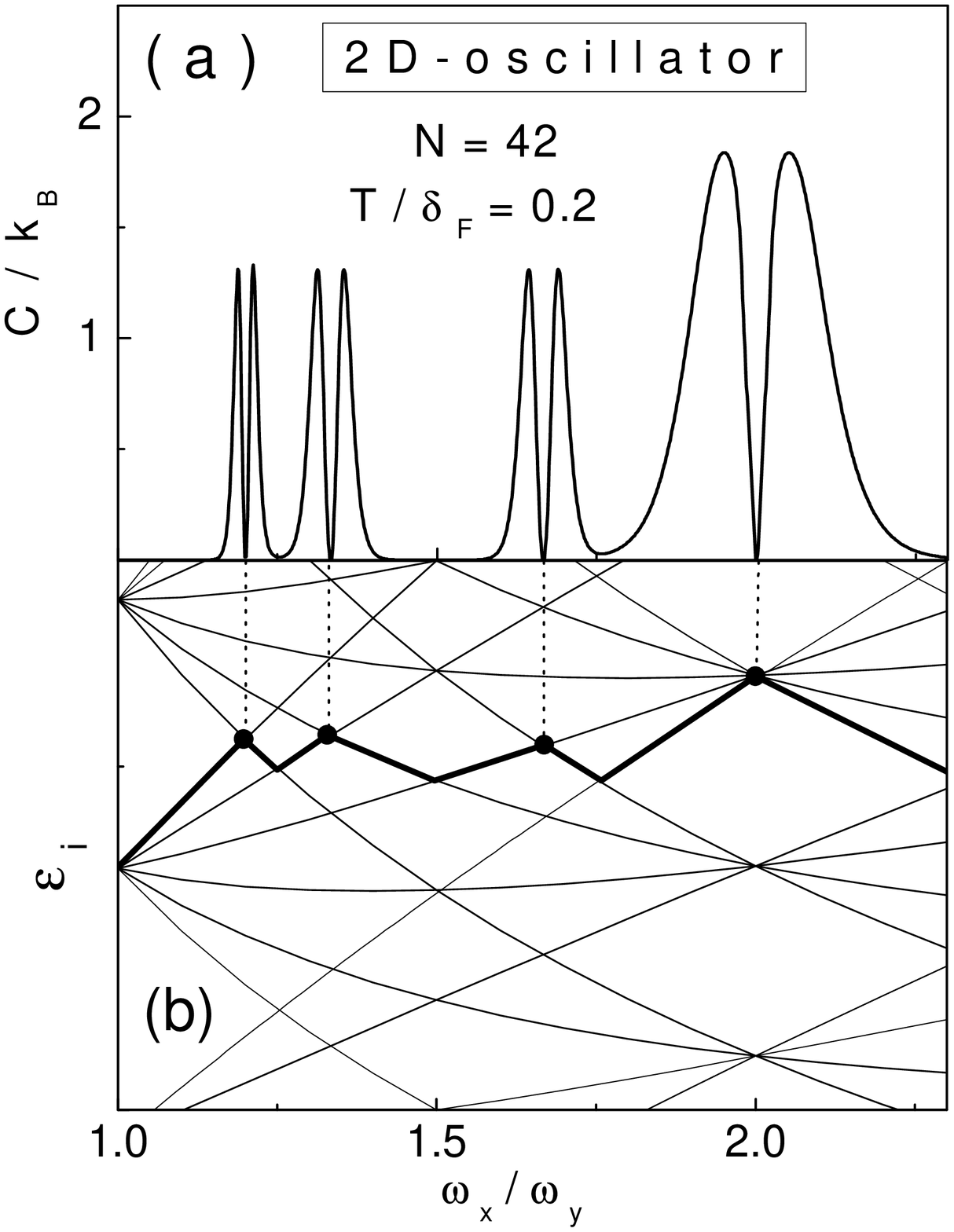}}
\caption{{\label{Lay42T012}}(a) - Shape oscillations of the low temperature canonical heat capacity of the system with $N=42$ fermions moving in $2D$ oscillator potential. $\omega_x$, $\omega_y$ are the oscillator frequencies. $\delta_F=\varepsilon_F/2N$.  (b) - Electron level energies near the Fermi level (bold line) vs $\omega_x/\omega_y$. Solid circles are the crossing points of the Fermi level and upper levels free from electrons at $T=0$.}
\end{figure}
In Ref.~\cite{kuzmenko1} we showed that similar oscillations in $C$ (and in the magnetic susceptibility) are stimulated by the increasing magnetic field $H$ i.e. in that case the role of $\alpha$ was performed by $H$.

In $NPM$ the level crossing ($\alpha <\alpha_0$) or splitting ($\alpha >\alpha_0$) in the whole many electron system at small deviations of $\alpha$ from  $\alpha_0$ are produced due to transformations in the single - electron spectrum: alterations in $\alpha$ cause first, at $\alpha <\alpha_0$, confluence of two or several levels in a high degenerated level ($\alpha =\alpha_0$) then, at $\alpha >\alpha_0$, they split this level into levels with lesser degeneracies. However, the oscillations in  $C$ created by deformations are possible only near such points $\alpha_0$ in which among crossing or splitted levels there are the Fermi level and some levels with higher energies that are free from  electrons at $T=0$. In other words, the high degenerated single-electron Fermi level arising after crossing at $\alpha =\alpha_0$ has to be incompletely filled at $T=0$ i.e. $n_F <d_F$.

The heat capacity being plotted vs $\alpha$ at a fixed small temperature ($T~<\Delta E_{sp}(\alpha )$) varies near point $\alpha_0$ like the spectral heat capacity distribution $\varphi(\varepsilon)$ vs $\varepsilon$ as on the curve of $C$ vs $\alpha$ there must be, as discussed above, two maxima to the left and right of point $\alpha_0$ in which $C\simeq 0$ if $\Delta E_{sp}(\alpha )<\Delta E$. $C$ as a function of $\alpha$ possesses the left-right symmetry with respect to point $\alpha_0$ if levels with identical degeneracies and equal spacing between them merge in $\alpha_0$ and go out of this point. This symmetry implies that the amplitudes and distances from point $\alpha_0$ are equal for both maxima in $C$. The crossing of two levels with different degeneracies leads to asymmetry in maxima. That maximum is higher which is formed by crossing the completely filled Fermi level ($n_F=d_F$) and empty level $F+1$. The higher maximum is further from the point $\alpha_0$ than the second lower maximum.

In both cases, just mentioned, the functional dependence of $C$ on $T$ and on $\delta$ (the gap between adjacent crossing levels) cannot be considered separately because $C$ is a function of the product $\beta\delta$. This quantity in the maximum $(\beta\delta)_m$ depends on the degeneracy of the joint level in point $\alpha_0$ and on its occupation number: for crossing two doubly degenerated levels with two electrons on the lower of them $(\beta\delta)_m\approx 3$ as it takes place for the two level model considered in the previous section. Consequently the position of the maximum in $C$ on the axis $\alpha$ depends on the temperature: if at a fixed temperature e.g. $T_1$ its position is $|\alpha_1 -\alpha_0|$ then at $T_2>T_1$ the maxima move aside from point $\alpha_0$ i.e. $|\alpha_2 -\alpha_0|>|\alpha_1 -\alpha_0|$  since $\delta$ linearly varies with $|\alpha -\alpha_0|$ provided the deviations are small. Whereas the positions of the minima in $C$ are strictly fastened by the positions of the points $\alpha_0$ where $(\beta\delta)_0\sim\beta(\alpha -\alpha_0)=0$. This link of $T$ and $\alpha$ implies that at each temperature (one case is e.g. given in Fig.~\ref{Lay42T012}) the deformation $|\alpha -\alpha_0|$ is tuned in so that the positions of the peaks in the spectral heat capacity  $\varphi(\varepsilon)$ coincide with the energies (depending on $\alpha$) of the crossing or splitted resonance levels. Therefore if it were possible to decrease the temperature in such a way that the ratio $T/|\alpha -\alpha_0|$ remained equal to a constant then maxima in $C$ would be observed almost in the immediate vicinity of $\alpha_0$.

To exemplify low temperature oscillations in $C$ initiated by the deformation we have calculated in the framework of $NPM$ the electron heat capacity of such systems as an elliptic $2D$-oscillator system with variable ellipse semiaxes and $3D$-system in a circular cylindrical rectangular potential with variable radius $R$ and height $H$. In both cases the spatial area or volume keeps a constant value at all shape variations i.e. the particle number remains invariable.

For $2D$-oscillator systems we have taken advantage of the selfconsistency condition suggested in Ref.~\cite{bohr}.  This condition consists in that the shape of the particle spatial distribution (in the $2D$-case it is $x^2a_x^{-2}+y^2a_y^{-2}$) has to be similar to the shape of the average potential in which particles move (for $2D$-oscillators it is $x^2\omega_x^{2}+y^2\omega_y^{2}$, $\omega_0=\omega_x\omega_y=const$, the latter is the spatial area conservation condition). Hence, the ellipse semiaxes $a_x$,  $a_y$ have to be proportional to reciprocal oscillator frequencies $\omega_x$, $\omega_y$, $a_x\sim\omega_x^{-1}$, $a_y\sim\omega_y^{-1}$, i.e. deformations of an ellipse can be defined via $2D$-oscillator frequencies. Therefore in this case we choose the ratio of $\omega_x/\omega_y$ as a deformation parameter $\alpha$.

The $2D$-oscillator eigenvalues can gain an additional degeneracy (as compared with the spin degeneracy) if $\alpha_0=\omega_x/\omega_y$ is equal to the ratio of integers $\nu_1/\nu_2$. The Fermi energy limits possible sets for $\nu_1$ and $\nu_2$, besides, as discussed above, the heat capacity oscillations arise only near such points $\alpha_0$ in which (at $T=0$) the Fermi level crosses levels free  from electrons. In Fig.~\ref{Lay42T012} this condition is fulfilled only in four points of $\alpha_0$ ($6/5$, $4/3$, $5/3$ and $2/1$). In these points the additional degeneracy of the Fermi level is $2$ for the first three values of $\alpha_0$ and $5$ for $\alpha_0=2$, as it shown in the bottom panel of Fig.~\ref{Lay42T012}.  At small deviations of  $\alpha$ from $\alpha_0=2$ all $5$ levels are practically equidistant.

\begin{figure}[p]
\scalebox{0.5}
{\includegraphics{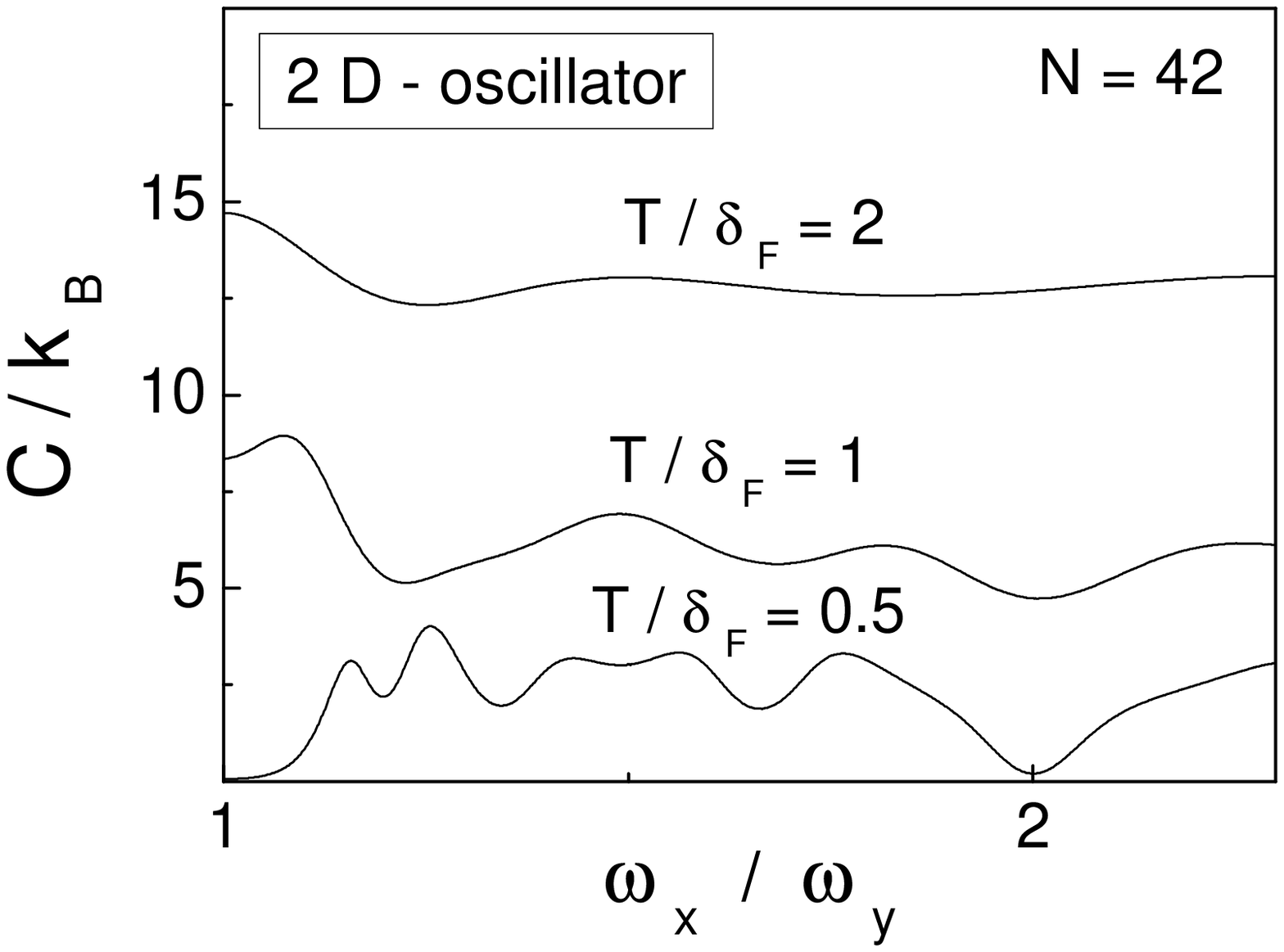}}
\caption{{\label{Lay2Dosc1952}} The temperature evolution of the shape oscillations of $C$ for the same system as in Fig.~\ref{Lay42T012}. $\omega_x$ and $\omega_y$ are the oscillator frequencies.}
\end{figure}

Near these points $\alpha_0$ i.e. before crossing and after splitting each level is twice (spin) degenerated and so the heat capacity vs $\alpha$ has mirror symmetry with respect to the perpendicular to point $\alpha_0$ and its maxima are determined practically only by values of $\beta\delta$. For the first three cases (near $\alpha_0=6/5, 4/3, 5/3$) the values of $\delta$, as it seen in Fig.~\ref{Lay42T012} under corresponding maxima, are nearly identical that results in the identical heights of their maxima. In the last case (near $\alpha_0=2$) $\delta$ is smaller than in previous cases and the maxima are higher, $C\sim (\beta\delta)^2\exp(-\beta\delta)$. The distances between maxima  are explained by the relationship between $\delta$ and $\Delta\alpha=|\alpha -\alpha_0|$. For small $\Delta\alpha$: $\delta\simeq\omega_0(\alpha_0)^{-1/2}\nu_2^2\Delta\alpha$, ($\alpha_0=\nu_1/\nu_2$). Since for all cases in Fig.~\ref{Lay42T012} values of $\delta$ differ insignificantly and the temperature is independent of $\alpha$ the greater distances between maxima correspond to the greater values of $\alpha_0$ (and thereby the lesser values of $\nu_2$), e.g. if $\alpha_0=4/3$, $\delta\simeq 2.6\omega_0\Delta\alpha$ but if $\alpha_0=2/1$, $\delta\simeq 0.7\omega_0\Delta\alpha$. Fig.~\ref{Lay2Dosc1952} shows the temperature evolution of the maxima. Rise in temperature leads first to confluence of the low temperature deformation maxima then weak extrema determined by spacings between shells can be revealed and at $T>2\delta_F$ all oscillations of $C$ vs $\alpha$ are smoothed.

In the previous section in Figs.~\ref{CClin3540} and \ref{Cube1008} values of $C$ vs $T$ are depicted for such symmetric systems  as spheres and cubes. Now Figs.~\ref{Lay42T012} and ~\ref{Lay2Dosc1952} can be used to explain how the variation of $C$ vs $T$ will change at symmetry breakdown. First we will consider two extreme cases: the high degenerated Fermi level is completely occupied at $T=0$ i.e. $n_F=d_F$ and the second case is $n_F\ll d_F$. The first case is realized in that point $\alpha$ in Fig.~\ref{Lay42T012} where $\alpha_0=(\omega_x/\omega_y)_0=1$ and $d_F=n_F=12$. Here small symmetry disturbance practically changes nothing: the maximum in $C$ appears at the resonance temperature $T\simeq 0.2(\varepsilon_{F+1}-\varepsilon_F)$, as it observed in Fig.~\ref{Lay2Dosc1952} at $T/\delta_F=2$. In the second case ($\omega_x/\omega_y=2$, $d_F=10$, $n_F=2$) in Fig.~\ref{Lay42T012} at small deformations $\Delta\alpha\simeq 0.05$ there arise two low temperature maxima whereas at the resonance temperature $T\simeq 0.2(\varepsilon_{F+1}-\varepsilon_F)$, where $\varepsilon_{F+1}$ and $\varepsilon_F$ are values at $\omega_x/\omega_y=2$, a maximum (or even a bend) on the curve of $C$ versus $T$ is practically unobserved both in the symmetric case $\omega_x/\omega_y=2$ and at small symmetry disturbance $\omega_x/\omega_y=2\pm 0.05$ (the condition $n_F\ll d_F$ is discussed in connection with Fig.~\ref{C2levd10}). In the intermediate cases ($n_F\sim d_F/2$) two types of local maxima have to occur: at $T$ nearly equal to $T_{res}$ of the symmetric system and at $T\ll T_{res}$ corresponding to small splitting of the Fermi level at deformation. Thus, small symmetry disturbances do not practically change the positions of those maxima in $C$ which occur in the symmetric case but at much lower temperatures there can appear other maxima if $n_F<d_F$.

\begin{figure}[p]
\scalebox{0.7}
{\includegraphics{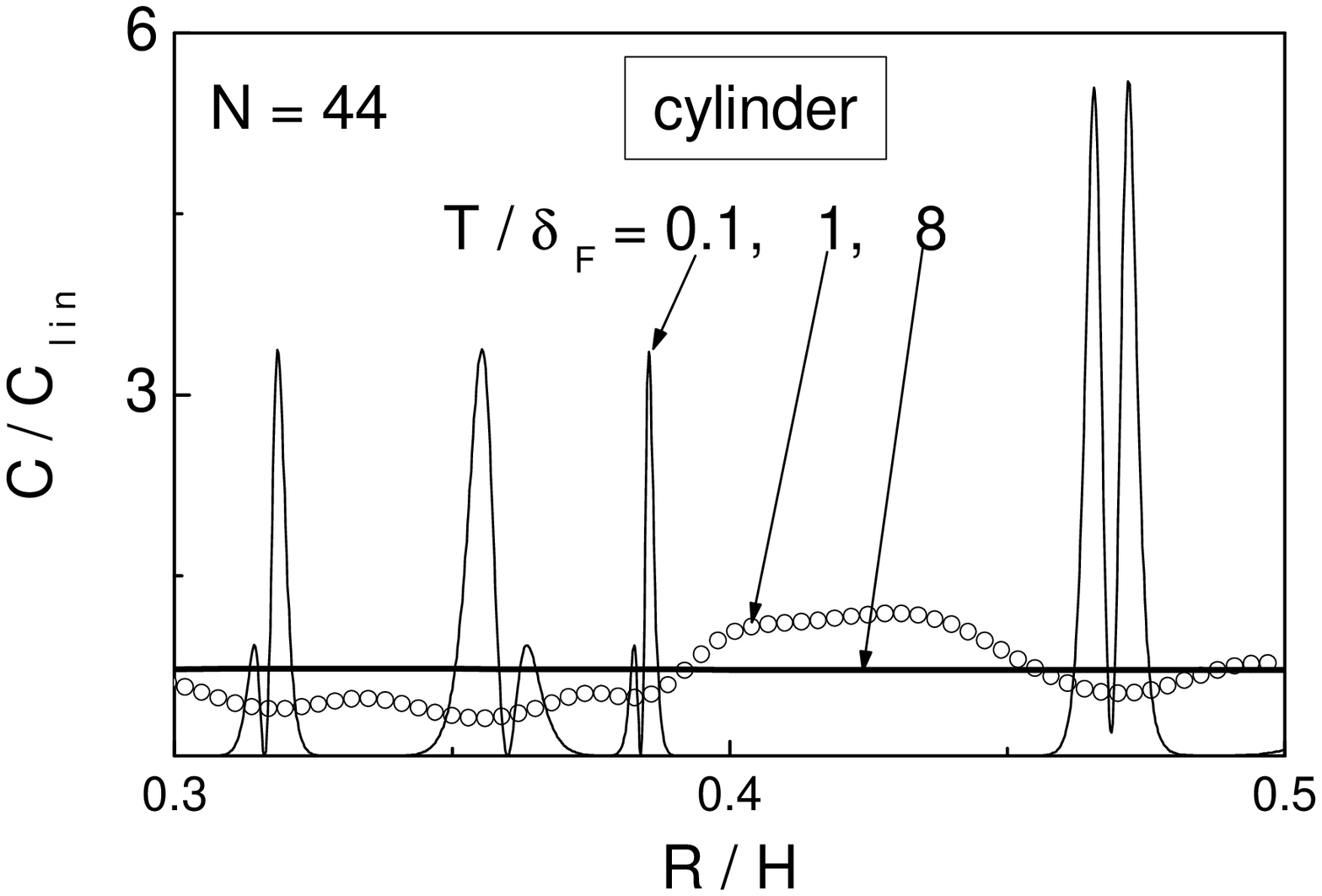}}
\caption{{\label{CylOsc}} Shape oscillations of the low temperature canonical heat capacity vs $R/H$ for the $N=44$ system in the $3D$ rectangular  circular symmetric cylindrical cavity with radius $R$ and height $H$.  $\delta_F=4\varepsilon_F/3N$.}
\end{figure}

At stretching (squeezing) a circular cylinder along its symmetry axis (axis $z$)  that is accompanied by decreasing (increasing) its cross-section owing to the volume conservation the appearance of deformation extrema in $C$ is caused by accidental level degeneracies at a proper relationship between radius $R$ and height $H$ of the cylinder. Their ratio is therefore adopted to be a deformation parameter $\alpha=R/H$. The degeneracy of single-particle levels in a potential with circular symmetry arises due to spin degree of freedom and in consequence of symmetry of the potential (conservation of angular momentum $z$-projection $\Lambda$) i.e. the degeneracy takes values equal to $4$ ($|\Lambda|\neq 0$) or $2$ ($\Lambda=0$). It means that levels merging in points $\alpha_0$ or going out of these points can possess different degeneracies. As pointed out above, this difference in degeneracies leads to asymmetry of the maxima disposed around points $\alpha_0$ on the heat capacity curve vs $\alpha$. Such cases are shown in Fig.~\ref{CylOsc} where each of the first three level crossing points ($C\simeq 0$ in these points) is bordered by two maxima with distinct amplitudes while the last case displays the symmetric picture. The following degeneracies and occupation numbers at $T=0$ can be ascribed to levels $F$ and $F+1$: near the first and third points $\alpha_0$ in Fig.~\ref{CylOsc} at crossing  $d_F=4$, $n_F=2$, $d_{F+1}=2$, $n_{F+1}=0$, at splitting $d_F=n_F=2$, $d_{F+1}=4$, $n_{F+1}=0$; near the second point $\alpha_0$ at crossing $d_F=n_F=4$, $d_{F+1}=2$, $n_{F+1}=0$, at splitting $d_F=n_F=2$, $d_{F+1}=4$, $n_{F+1}=2$ and near the last point $\alpha_0$ given in Fig.~\ref{CylOsc} at crossing and splitting $d_F=n_F=d_{F+1}=4$, $n_{F+1}=0$.
Thus, in all points $\alpha_0$ (when levels $F$ and $F+1$ have merged) the high degenerated level ($d=d_F+d_{F+1}$) is unfilled otherwise the maxima in $C$ near points $\alpha_0$ would be impossible. The adduced numbers of $d$ and $n$ explain the values of amplitudes in maxima in accordance with the rule given above which can be detailed for two crossing levels $F$ and $F+1$ if one writes out the leading term in the equation for $C$:
\begin{displaymath}
C\sim n_F(d_{F+1}-n_F)\left [(n_{F+1})(d_F-n_F+1)\right ]^{-1} (\beta\delta )^2\exp(-\beta\delta),
\end{displaymath}
where $\delta$, as before, is the difference $\varepsilon_{F+1}-\varepsilon_F$. The temperature evolution of the extrema in $C$ shown in Fig.~\ref{CylOsc} develops approximately in the same manner as in previous case in Fig.~\ref{Lay42T012}.

The examples of the heat capacity shape dependence considered in this section indicate that in spite of the differences in space dimensionality and conditions governing the appearance of the heat capacity extrema the shape of the $N$ particle system can be adjusted so (most likely it is possible to make only experimentally) that in a low temperature range $T<\varepsilon_F/N$ the fermion heat capacity can attain to its maximum or any lesser value at a given $N$ (the needed value of $C$ depends on the problem where these values are of particular importance). This property of mesoscopic systems can reveal itself also in granular systems.

\begin{figure}[p]
\scalebox{0.7}
{\includegraphics{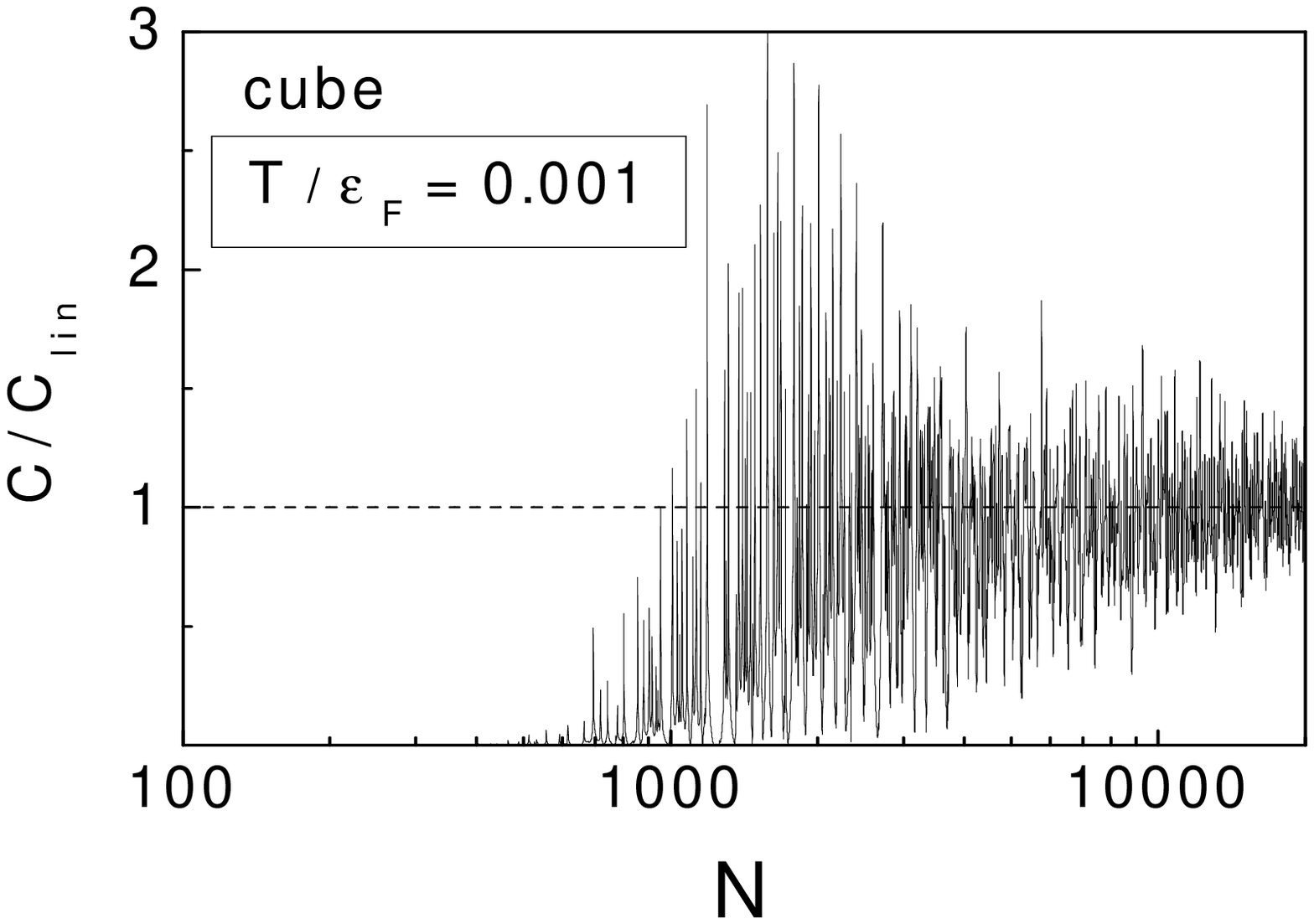}}
\caption{\label{Rubik}  $N$- oscillations of the canonical heat capacity ($C/C_{lin}$) in cubes, $N\leq 3\cdot 10^4$.}
\end{figure}

To show it consider two mesoscopic systems. The first is a granular system which is composed of close-packed metallic ultrasmall granules as it is described e.g. in Ref.~\cite{beloborodov}. These granules are assumed to be in thermal contact i.e. they are kept at common temperature but the mutual electron exchange is impossible (the very weak intergranular electron coupling). Here we again idealize the situation and suppose that each granule contains $N$ conduction electrons, all they are identical in size and shape and the heat capacity of each granule at a fixed low temperature ($<\varepsilon_F/N$) is equal to $C(N)\neq C_{lin}(N)$, Eq.~(\ref{Clin1}). If the granular system consists of $n$ such granules (the summary electron number $N_s=nN$) its heat capacity, $nC(N)$, can be essentially distinct from $C$ of the second system which is a relatively large monolithic mesoscopic system with the same number of conduction electrons $N_s$. Here the Fermi energies of each granule and the monolithic system are assumed to be practically the same. Since the mean level spacing in the monolithic system is $n$ times smaller than in the granule the fixed temperature can appear so high for the monolithic system at $n\gg 1$ that its heat capacity can take only the shape-independent Sommerfeld's value $C_{lin}(N_s)=nC_{lin}(N)$. Therefore the ratio of the electron heat capacities of the granular system and the monolithic one is equal to the quantity $C(N)/C_{lin}(N)$ given in some figures of   this and previous sections such as Figs.~\ref{Br14391442},~\ref{Cube1008},~\ref{2maxIntFunSph},~\ref{2maxIntFunBr}. Thus these figures gain the second content: they give not only values of $C/C_{lin}$ vs $T$ for a system with a fixed particle number $N$ but the same curves describe the ratios of $C$ in granular and monolithic systems, where the former are composed of small $N$ particle granules: briquettes (Figs.~\ref{Br14391442},~\ref{2maxIntFunBr}), cubes (Fig.~\ref{Cube1008}) and spheres (Fig.~\ref{2maxIntFunSph}). Additionally in Fig.~\ref{Rubik} we show the variety of values  $C/C_{lin}$ for cubic systems in a wide range of $N\leq 3\cdot 10^4$.

\section{Conclusion}

For description of fermion properties of mesoscopic $N$-particle systems which are in thermal equilibrium with surroundings (the canonical ensemble) in the framework of $NPM$ we have suggested a new formalism representing the single-particle level occupations as a continuous function of energy $\varepsilon$ similar to the Fermi-Dirac distribution. In $NPM$ many static thermodynamic quantities can be obtained by applying this representation. In particular it allows the electron heat capacity $C$ to be calculated as a convolution of the state density with function $\varphi(\varepsilon)$, the spectral distribution of $C$ that gives the contribution of each single-electron state to $C$. The two hamped shape of $\varphi(\varepsilon)$ vs $\varepsilon$ gives possibility to account for the appearance of local maxima in $C$ within the low temperature range ($T<\delta_F$, the latter is the mean level spacing near $\varepsilon_F$) by virtue of the resonance amplification of contributions of states nearest $\varepsilon_F$ if the positions of two peaks in $\varphi(\varepsilon)$ coincide with energies of either two high degenerated levels near $\varepsilon_F$ (or level bunchings) or even two solitary double degenerated levels. However $C$ vs $T$ for the uniform spectrum with equal spacing between spin degenerated levels does not reveal noticeable maxima. It is worth mentioning that local maxima in $C$ are observed more distinctly on curves $C/C_{lin}$ (or $C/T$) vs $T$ where $C_{lin}$ is the Sommerfeld's electron heat capacity linear in $T$.

The temperature of the maximum in $C$ (the resonance temperature $T_{res}$) makes up a part of $\Delta\varepsilon$, the difference between energy levels in the vicinity of $\varepsilon_F$ and depends on $n_F$, the Fermi level occupation number at $T=0$. So if $n_F=d_F$ is high ($d_F>2$) ($d_F$ is the Fermi level degeneracy) then $T_{res}\simeq 0.2(\varepsilon_{F+1} -\varepsilon_F)$ whereas for the case of two solitary spin degenerated levels ($\varepsilon_F$ and $\varepsilon_{F+1}$)  with $n_F=1$ this temperature is equal to  $\simeq 0.4(\varepsilon_{F+1} -\varepsilon_F)$. Thus,  establishing $T_{res}$ and studying $N$-dependence of $C$ in samples of identical shapes can give information concerning the single-electron level structure near $\varepsilon_F$.

Within the low temperature range the heat capacity appears to be shape-dependent at an invariable spatial area (for $2D$-systems ) or volume (for $3D$-systems). At continuous alteration of some parameter $\alpha$ characterizing deformation of the system, $C$ can display oscillations vs $\alpha$ with minimum values of $C$ at such deformations where the system gains an additional symmetry.

It is shown that at $T<\varepsilon_F/N$ the ratio of the electron heat capacities of a granular system consisting of $n$ identical insulted $N$-electron subsystems ($n\gg 1$) and a monolithic system with the same summary electron number ($n\cdot~N$) is equal to $C(N)/C_{lin}(N)$ which can be stretched from zero to several units.

On the whole, the presented material can be viewed as evidence of possible essential deviations of low temperature ($T<\delta_F$) properties of electron mesoscopic systems from those established for $T>\delta_F$ that is caused by the irregularities in single-electron spectra and this has to be taken into account at consideration of both the heat capacity and all other thermodynamic properties of finite electron systems.

\ack
 We thank V. Z. Kresin, F. Philippe, P. Hawrylak and O. Bourgeois for useful remarks. This work is supported by the ISTC under grant Nr.~3492.

\end{document}